\documentclass{webofc}

\newcommand{\DLR}{\smash{\overset{\text{\small$\leftrightarrow$}}{\smash{D}\vphantom{+}}}}

\newcommand{\bea}{\begin{eqnarray}}
\newcommand{\eea}{\end{eqnarray}}

\usepackage[varg]{txfonts}   
\usepackage{subcaption}
\usepackage[pscoord]{eso-pic}

\newcommand{\placetextbox}[3]{
  \setbox0=\hbox{#3}
  \AddToShipoutPictureFG*{
    \put(\LenToUnit{#1\paperwidth},\LenToUnit{#2\paperheight}){\vtop{{\null}\makebox[0pt][c]{#3}}}
  }
}

\begin{document}

\placetextbox{0.82}{0.9}{WUB/22-04}
\title{Round table on Standard Model Anomalies}
%
%

\author{\firstname{Ashutosh V.} \lastname{Kotwal}\inst{1}\fnsep\thanks{\email{ashutosh.kotwal@duke.edu}} \and       \firstname{Joaquim} \lastname{Matias}\inst{2}\fnsep\thanks{\email{matias@ifae.es}}
             \and
      \firstname{Andrea} \lastname{Mauri}
      \inst{4,6}\fnsep\thanks{\email{a.mauri@cern.ch}}
      \and        \firstname{Tom} \lastname{Tong}
      \inst{5}\fnsep\thanks{\email{Tiantian.Tong@uni-siegen.de}}
             \and  
     \firstname{Lukas} \lastname{Varnhorst}\inst{3}\fnsep\thanks{\email{varnhorst@uni-wuppertal.de}}
}

\institute{Department of Physics, Duke University, Durham, NC 27708, USA 
\and
           Universitat Aut\`onoma de Barcelona, 08193 Bellaterra, Barcelona,
Institut de F\'{i}sica d’Altes Energies (IFAE), The Barcelona Institute of Science and Technology, Campus UAB, 08193 Bellaterra (Barcelona)
\and
           Department of Physics, University of Wuppertal, D-42119 Wuppertal, Germany
\and
           Imperial College of London, United Kingdom
\and
           Department of Physics, University of Siegen, D-57068 Siegen, Germany
\and
            Nikhef National Institute for Subatomic Physics, The Netherlands
          }

\abstract{%
  This contribution to ``The XVth Quark confinement and the Hadron spectrum conference"   covers a description, both theoretical and experimental, of the present status  of a set of very different anomalies. The discussion ranges from the long standing $b\to s\ell\ell$ anomalies, $(g-2)$ and the new $M_W$ anomaly. 
}
\maketitle
\section{Introduction}
\label{intro}

In recent years indirect New Physics searches have experienced a flourishing period. The observation of persistent (at least up to now) anomalies in a large set of observables has opened the window to what possibly are the first hints of New Physics.
Still further experimental scrutiny is necessary to confirm or dismiss some of the observed anomalies. We will refer to an anomaly generically as any observable exhibiting a ``significant discrepancy" between SM prediction and data.

 In the round table at the Confinement XV conference we discussed three different categories of anomalies 
 from a theoretical and experimental point of view. Therefore, we will organize  this proceeding accordingly. The three types of anomalies are: $b\to s\ell\ell$ anomalies, $g-2$ and the recent $W$ mass one that will be discussed in the following sections.

\section{$b \to s \ell\ell$ anomalies: Experiment}

This section presents a short review of current $b$-decays anomalies; existing measurements are grouped in different classes,
depending on the accuracy of their theoretical prediction.

\subsection{Lepton Flavour Universal observables and $B^0_{(s)} \to \mu\mu$ decays}

\subsubsection{$B_s \to \mu^+ \mu^-$ branching ratio}

\begin{figure}[t]  
\centering
\includegraphics[width=0.5\textwidth]{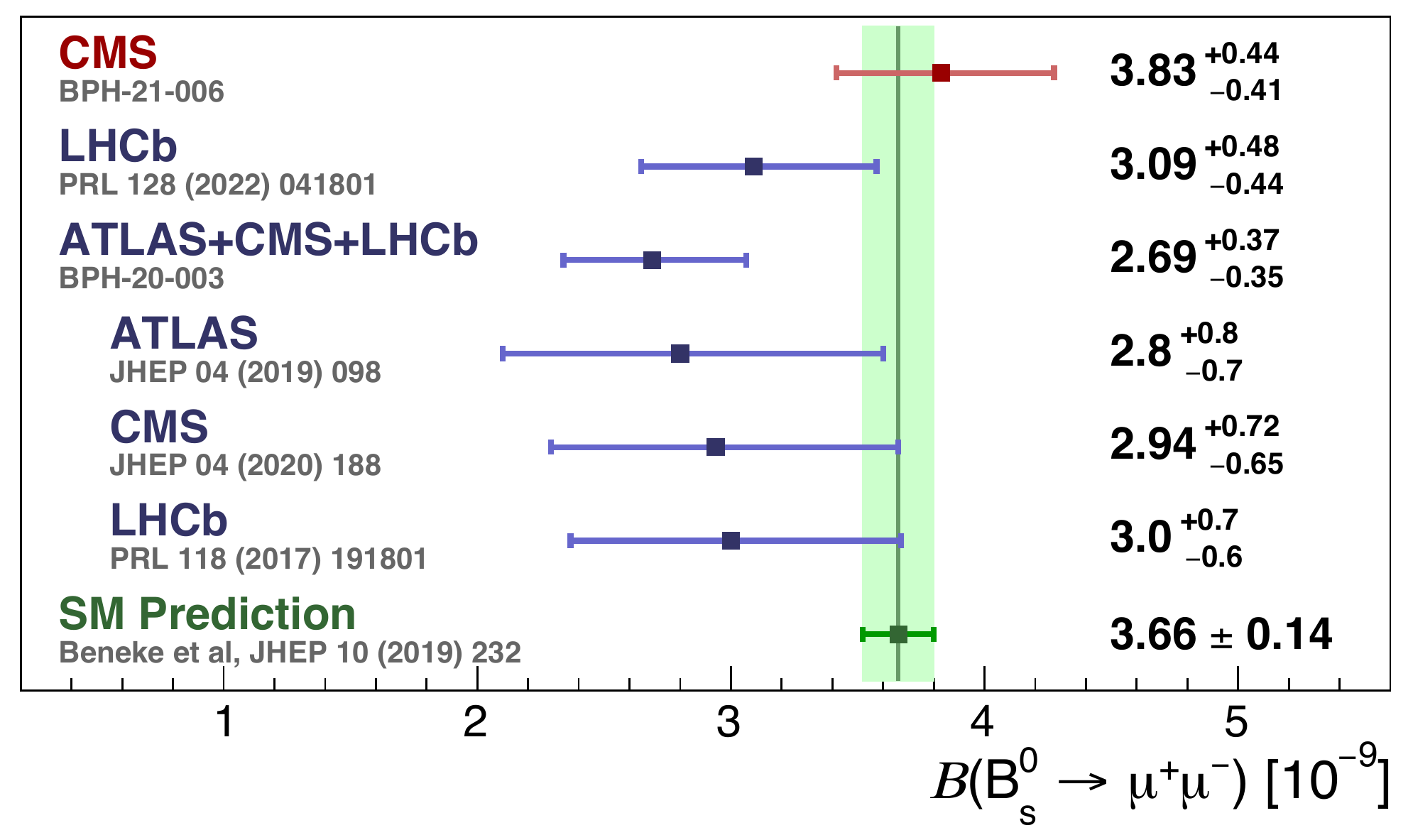}
\caption{Summary of all experimental measurements on $B_s \to \mu^+ \mu^-$ branching ratio.}
\label{fig:Bs2mumu}
\end{figure}

Decays of $B_s \to \mu^+ \mu^-$ are extremely rare in the Standard Model (SM), with a branching fraction of the order of
$10^{-9}$, since they do not only proceed through
loop diagrams but are also helicity suppressed.
They are considered a golden channel in flavour physics since their branching fraction can be calculated very precisely, 
with an uncertainty of the order of few percent~
\cite{Beneke:2019slt,PhysRevD.97.053007}.
$B_s \to \mu^+ \mu^-$ decays have been searched for decades till the first observation published by the LHCb experiment in 2012~\cite{LHCb:2012skj}.
Since then, more and more data have been collected and several updates of the same measurement have been released,
making our knowledge on the branching ratio of this decay more and more precise.
Fig.~\ref{fig:Bs2mumu} summarises all the currently available measurements on the $B_s \to \mu^+ \mu^-$ branching fraction, provided by 
the ATLAS~\cite{ATLAS:2018cur}, CMS~\cite{CMS:2019bbr,CMS-PAS-BPH-21-006} and LHCb~\cite{LHCb:2017rmj,LHCb:2021vsc} experiments,
as well as a partial combination of the different results from the three experiments~\cite{CMS-PAS-BPH-20-003}.
While during the last couple of years all the measurements appeared to prefer values that were lower than the SM prediction, 
the latest update from the CMS experiment~\cite{CMS-PAS-BPH-21-006} found a central value in line with the Standard Model:
 more data will certainly help in clarifying the picture.
Finally, a second global combination of the results from the three big collaborations
on the line of \cite{CMS-PAS-BPH-20-003} is expected once the update from ATLAS with the full Run~1 and Run~2 collected dataset will be released.

Moreover, with the increasing dataset, measurements of the $B_s \to \mu^+ \mu^-$  effective lifetime
have become possible.
While the experimental uncertainty is still large, $\mathcal{O}(10\%)$~\cite{LHCb:2021vsc,CMS-PAS-BPH-21-006}, 
one can expect in the future these observables to provide strong complementary constraints on New Physics models.
Finally, beyond the focus on $B_s \to \mu^+ \mu^-$ decays, these analyses also put a limit on the $B^0 \to \mu^+ \mu^-$ branching fraction,
whose ratio with respect to $B_s \to \mu^+ \mu^-$ represents an extremely clean test of the minimal flavour violation hypothesis,
as well as the $B_s \to \mu^+ \mu^- \gamma$ branching fraction.
Both limits are found to be of the order of 
$2\cdot 10^{-10}$~\cite{LHCb:2021vsc,CMS-PAS-BPH-21-006} and  $2\cdot 10^{-9}$~\cite{LHCb:2021vsc}, respectively.

\subsubsection{Lepton Flavour Universality test}

In recent years, great attention from the community has been focused on a class of measurements that showed some deviations with respect to the Standard Model predictions,
\textit{i.e.} Lepton Flavour Universality (LFU) tests.
These consist in measuring the ratio of branching fractions of the same decay process  with two different lepton flavours in the final state, namely
\begin{equation}\label{eq:RX}
R_X = \frac{\mathcal{B}( B \to X_s \mu^+ \mu^-)}{\mathcal{B}( B \to X_s e^+ e^-)} 
\end{equation}
where $X_s$ is a generic hadron containing a strange quark and muons or electrons as  lepton flavour final state.
This ratio can then be measured in different regions of $q^2$, the di-lepton invariant mass squared, and compared to the Standard Model hypothesis, 
which predicts this ratio to be equal to unity (in all cases where the mass of the leptons can be neglected) with very good precision~\cite{Bordone:2016gaq}.
In practise, when performing this measurement in the experiment, it is convenient to define instead the double ratio
\begin{equation}
R_X = \frac{\mathcal{B}( B \to X_s \mu^+ \mu^-)}{\mathcal{B}( B \to X_s e^+ e^-)} \cdot \frac{\mathcal{B}( B \to X_s J/\psi(\to e^+ e^-))}{\mathcal{B}( B \to X_s J/\psi(\to \mu^+ \mu^-))} 
\end{equation}
which profits from the use of the high statistics control channel $B \to X_s J/\psi(\to \ell^+ \ell^-)$ and allows the cancellation of the systematic uncertainties to a large extent.

\begin{figure}[t]  
\centering
\begin{subfigure}[b]{0.5\textwidth}
         \centering
         \includegraphics[width=\textwidth]{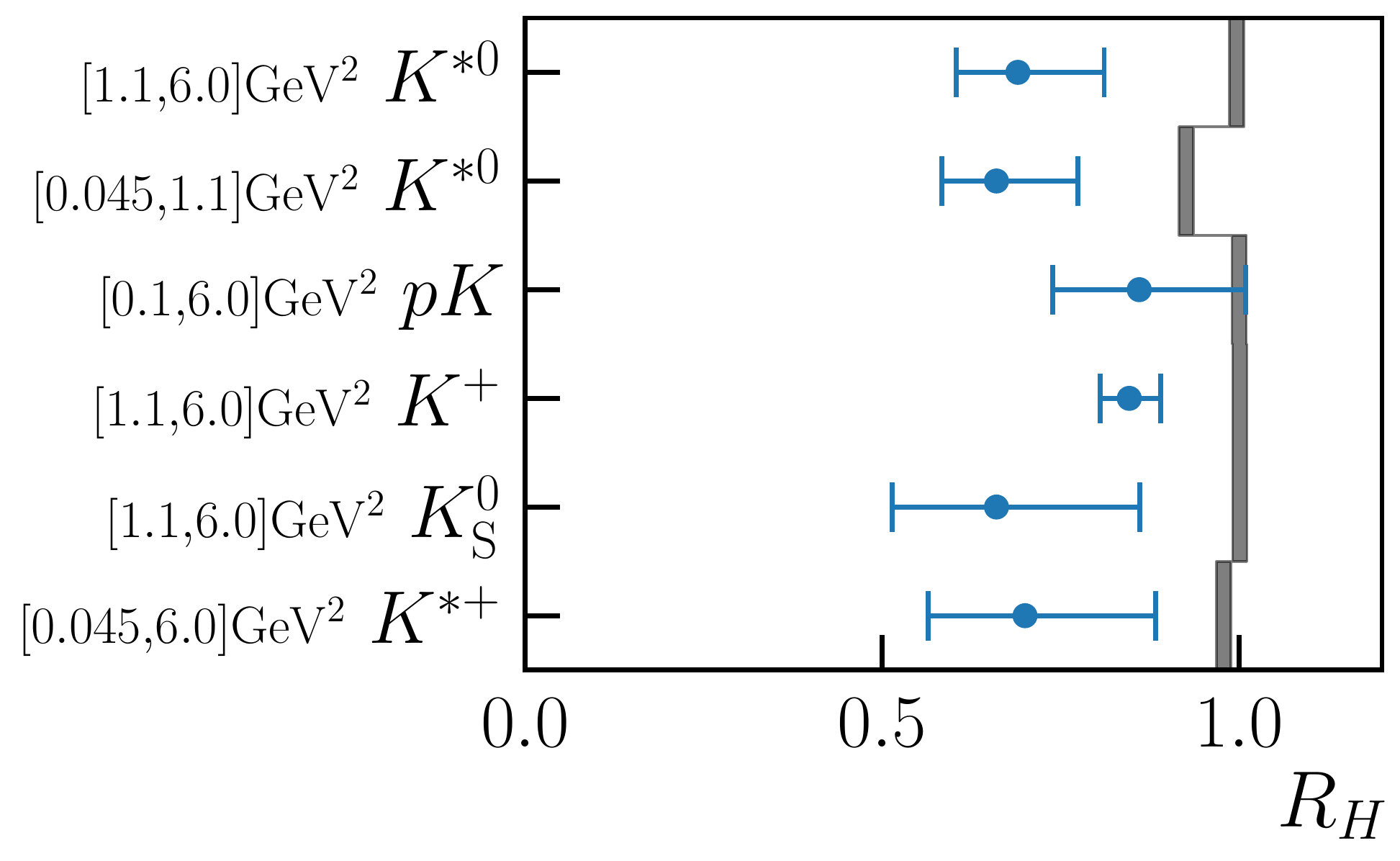}
\end{subfigure}
     \hfill
\begin{subfigure}[b]{0.4\textwidth}
         \centering
         \includegraphics[width=\textwidth]{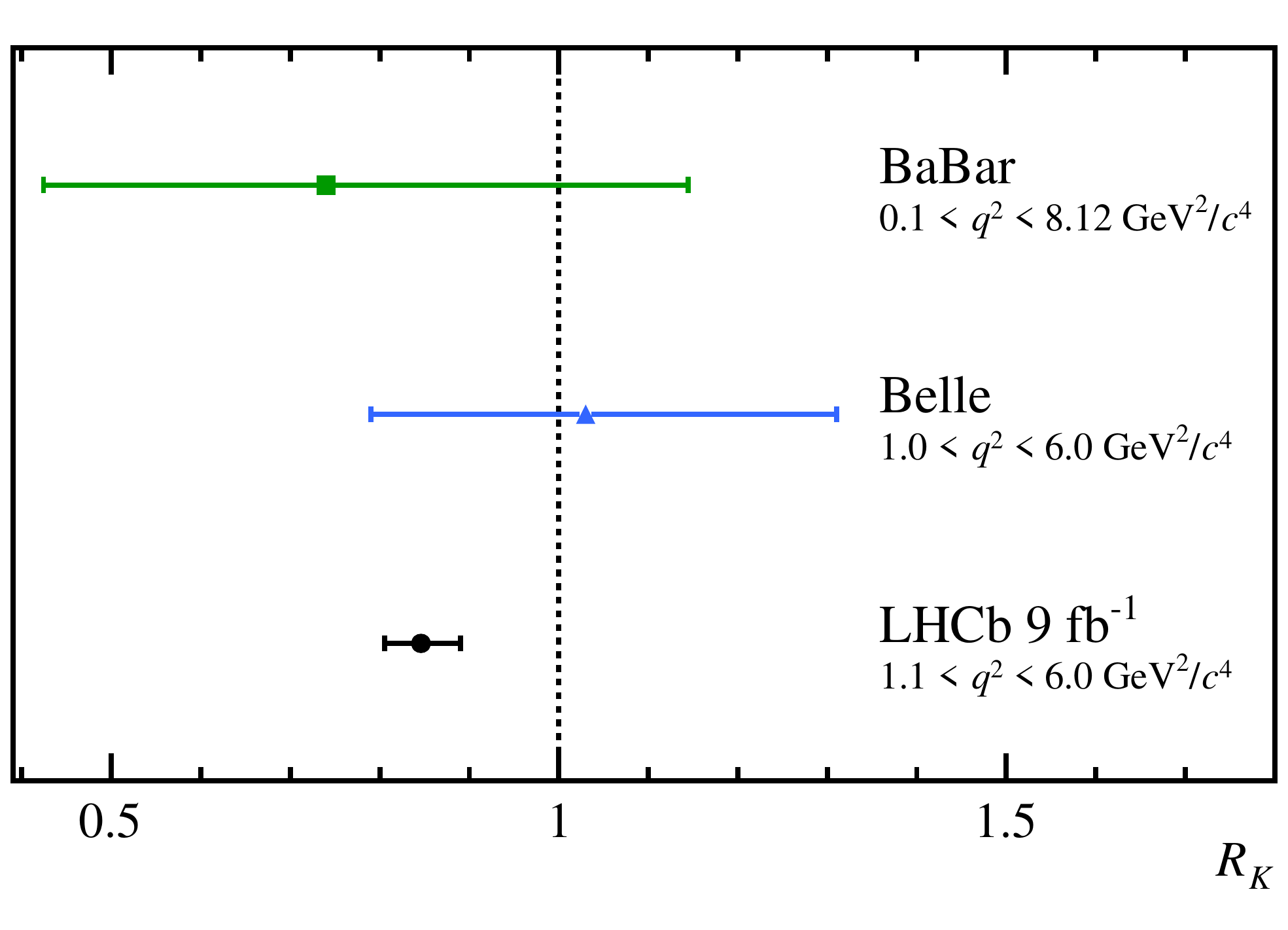}
         \vspace{-1mm}
\end{subfigure}
\caption{Left: summary of all LFU ratios measured in $b \to s \ell \ell$ decays published by the LHCb experiment. 
Right: the value of $R_K$ observed by the LHCb experiment~\cite{LHCb:2021trn} compared to the Babar~\cite{BaBar:2012mrf}  and Belle~\cite{BELLE:2019xld} measurements.}
\label{fig:RX}
\end{figure}

Following this strategy, the LHCb experiment measured such LFU ratios in several decay channels: 
$B^+ \to K^+ \ell \ell$~\cite{LHCb:2021trn}, 
$B^0 \to K^{*0} \ell \ell$~\cite{LHCb:2017avl}, 
$B^0 \to K_S^0 \ell \ell$~\cite{LHCb:2021lvy} and
$\Lambda_b \to p K \ell \ell$~\cite{ LHCb:2019efc}  
decays, whose results are schematically collected in Fig.~\ref{fig:RX}.
In general, all these measurements are found to be always below (but compatible with)  the SM prediction,
with the most precise being $R_K$, which is also the one exhibiting the largest tension with the SM.  
Given the importance of these anomalies and the difficulty from other experiments to provide competitive measurements 
(Fig.~\ref{fig:RX} right shows the value of $R_K$ obtained from the LHCb experiment compared to the other existing measurements 
from BaBar~\cite{BaBar:2012mrf} and Belle~\cite{BELLE:2019xld}), 
it will be crucial to clarify the origin of the observed deviation in the coming future.
Since all these measurements are statistically dominated, we can expect already the Run~3 of the LHC program to be able to refine the experimental situation.

\subsection{$b \to s \mu\mu$ observables}

\subsubsection{$b \to s \mu^+ \mu^-$ branching fraction}

\begin{figure}[t]  
\centering
\includegraphics[width=0.49\textwidth]{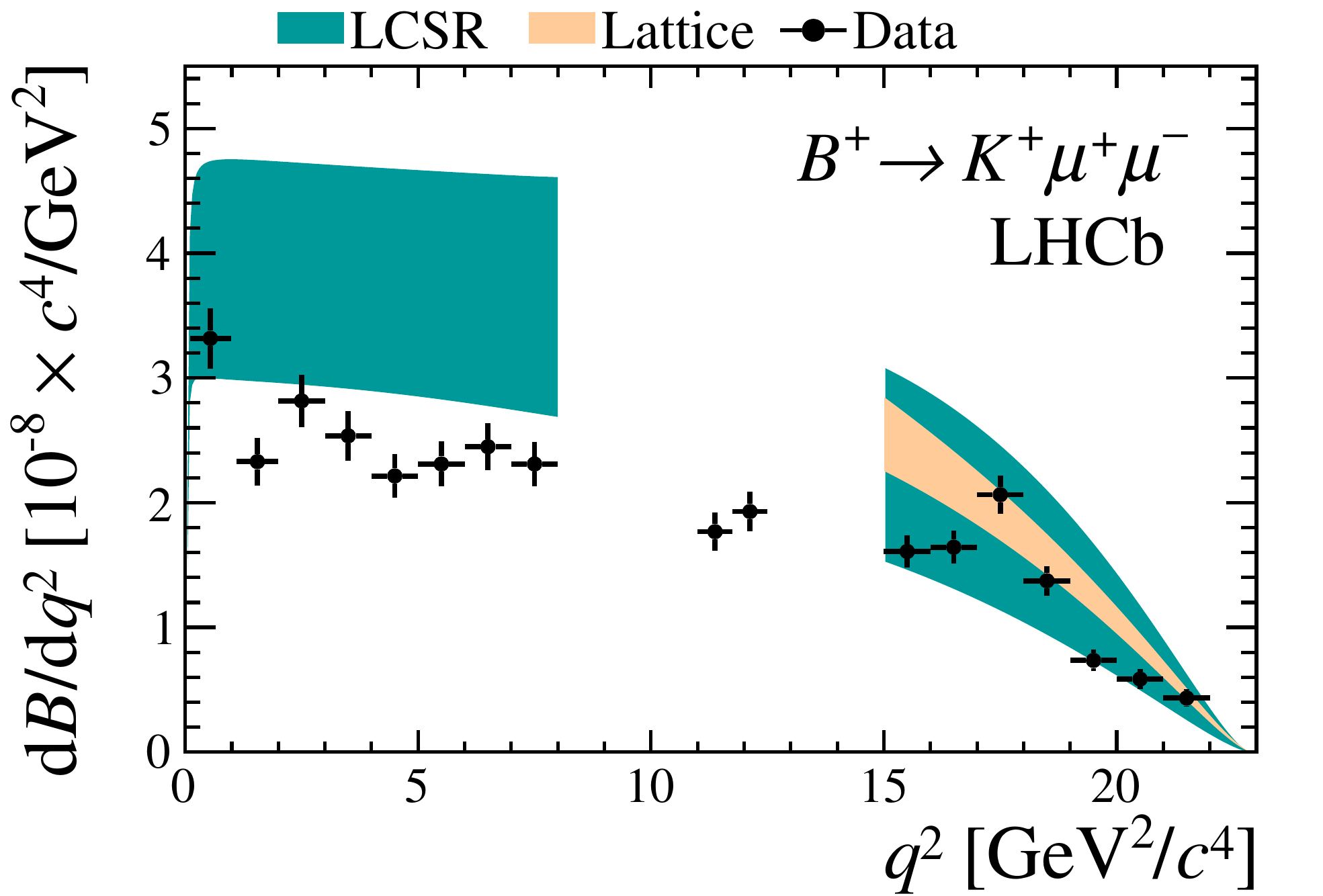}
\includegraphics[width=0.49\textwidth]{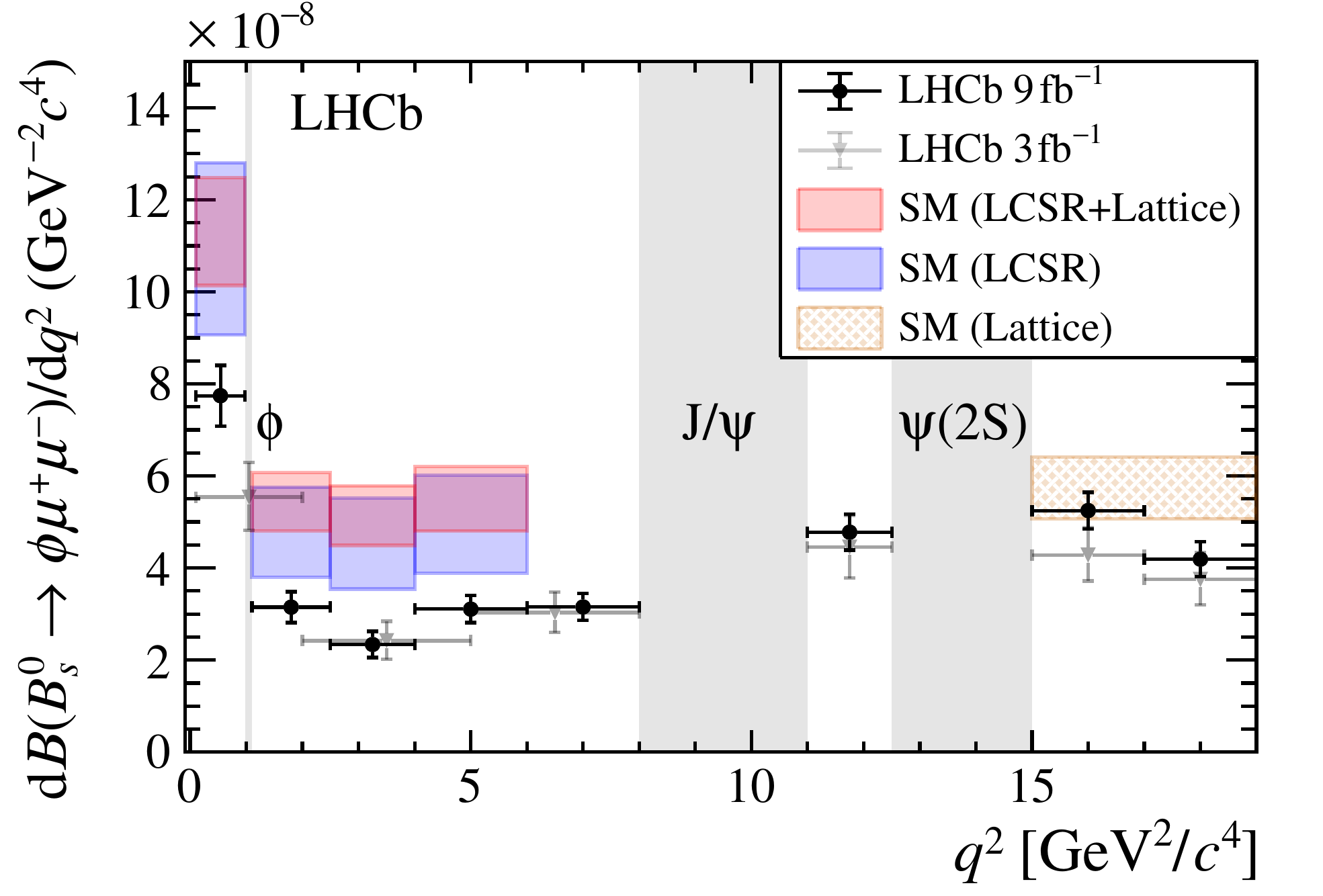}
\caption{Branching fraction measurements for (left) $B^+ \to K^+ \mu^+ \mu^-$~\cite{LHCb:2014cxe} 
and (right) $B_s \to \phi \mu^+ \mu^-$~\cite{LHCb:2021zwz} decays published by the LHCb experiment compared to the SM theoretical prediction.}
\label{fig:b2smm_BF}
\end{figure}

Besides double ratios, one can consider standard $b \to s \mu^+ \mu^-$ branching fraction measurements.
From the experimental point of view, decays with muons in the final state are easier to be analysed compared to their electron counterpart,
as they can profit from  higher selection efficiency, better resolution and particle identification, etc., however, from a theoretical point of view, 
 $B \to X_s \mu^+ \mu^-$ branching fractions suffer from large hadronic uncertainties.
 The LHCb experiment measured these branching fractions in a variety of different 
 final states~\cite{LHCb:2014cxe, LHCb:2021zwz, LHCb:2016ykl, LHCb:2015tgy};
 for illustration, the results obtained for $B^+ \to K^+ \mu^+ \mu^-$ and $B_s \to \phi \mu^+ \mu^-$ decays are reported in Fig.~\ref{fig:b2smm_BF}.
It is found that the experimental points always lie below the SM prediction, despite the large theoretical uncertainty.

\subsubsection{$B^{(+)} \to K^{*(+)} \mu^+ \mu^-$ angular observables}

\begin{figure}[t]  
\centering
\includegraphics[width=0.51\textwidth]{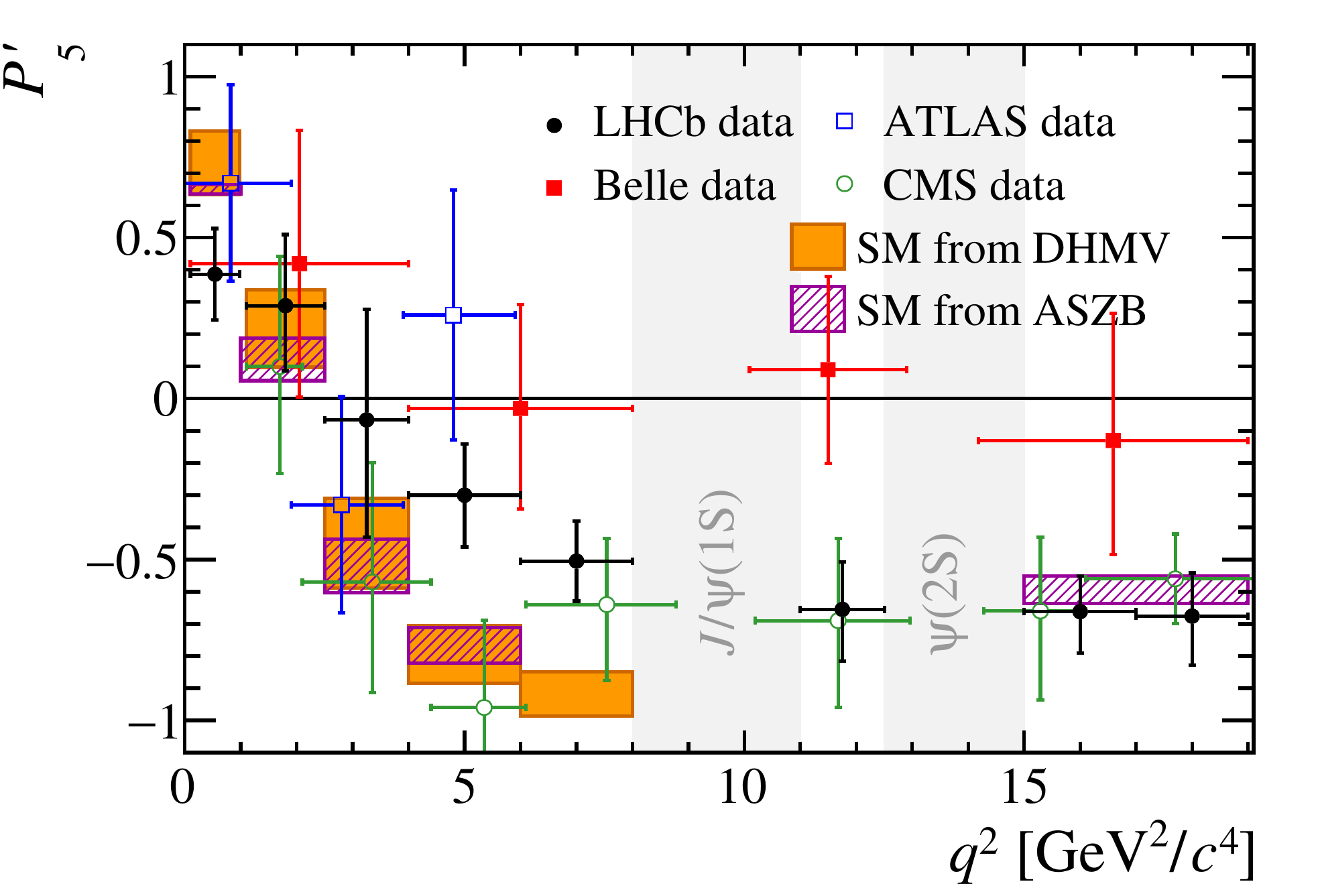}
\includegraphics[width=0.48\textwidth]{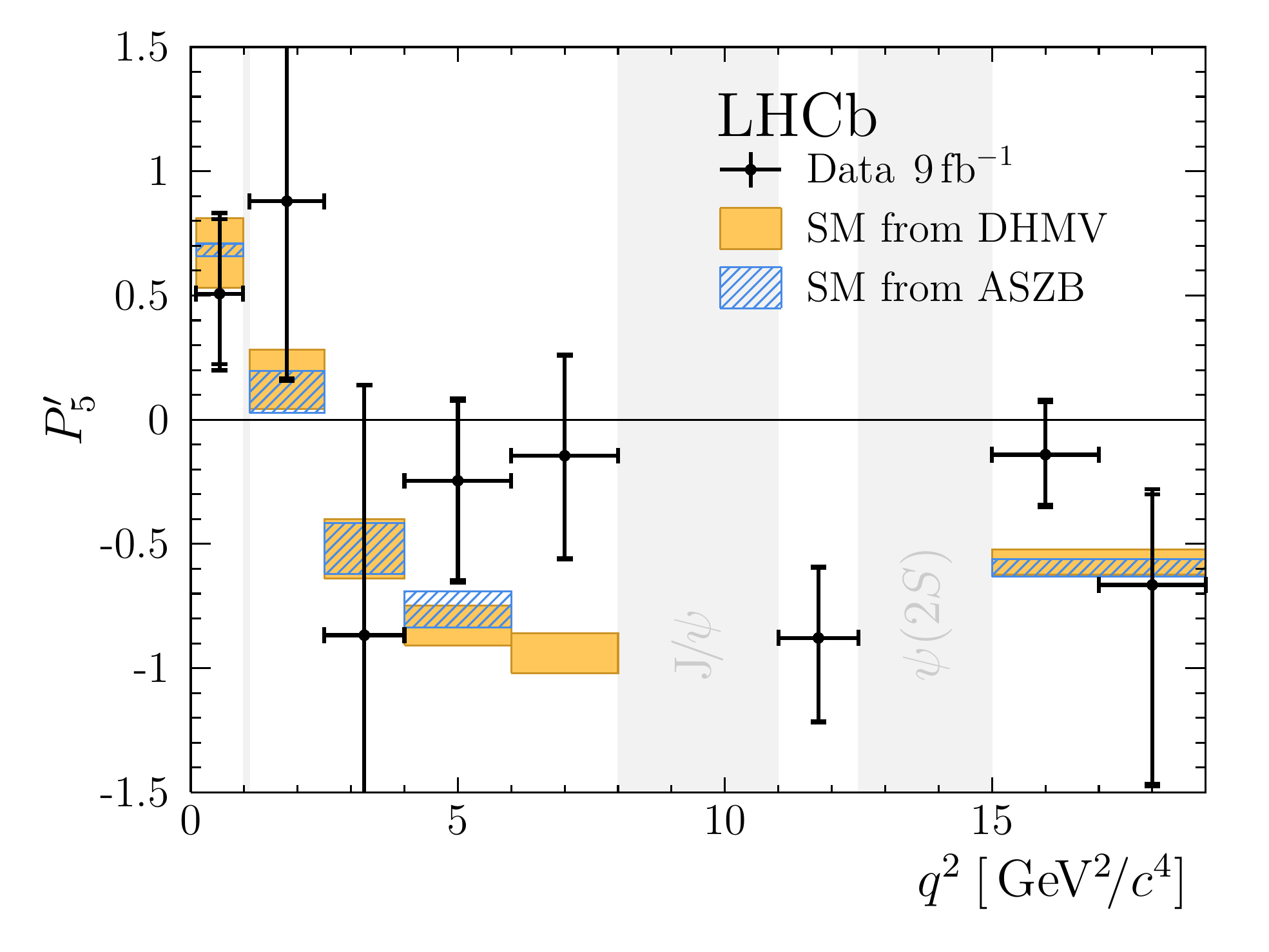}
\caption{Left: all different measurements of the $P_5^\prime$ angular observable in $B^0 \to K^{*0} \mu^+ \mu^-$ decays~\cite{LHCb:2020lmf,Wehle:2016yoi,ATLAS-CONF-2017-023,CMS-PAS-BPH-15-008}.
Right: measurement of $P_5^\prime$ provided by the LHCb experiment in $B^{+} \to K^{*+} \mu^+ \mu^-$ decays~\cite{LHCb:2020gog}.
In both cases, the SM theory predictions from two different groups ASZB~\cite{Bharucha:2015bzk,Altmannshofer:2014rta} 
and DHMV~\cite{Descotes-Genon:2014uoa} are overlaid for comparison.}
\label{fig:P5p}
\end{figure}

Thanks to the presence of a vector meson in the final state, $B^{(+)} \to K^{*(+)} \mu^+ \mu^-$ decays offer a great variety of angular observables to be studied.
The angular distributions of the final state particles, in fact, are highly sensitive to New Physics contributions.
In particular, the so-called $P_5^\prime$ observables attracted increasing interest from inside and outside the field thanks to its reduced 
theoretical uncertainty~\cite{Matias:2012xw,Descotes-Genon:2012isb} and a persistent set of measurements that points to a possible deviation with respect to the 
Standard Model.
Figure~\ref{fig:P5p} collects all the available $P_5^\prime$ measurement in 
$B^0 \to K^{*0} \mu^+ \mu^-$~\cite{LHCb:2020lmf,Wehle:2016yoi,ATLAS-CONF-2017-023,CMS-PAS-BPH-15-008} 
and $B^{+} \to K^{*+} \mu^+ \mu^-$~\cite{LHCb:2020gog} decays.
The experimental points are found to be higher than the SM prediction, especially in the $q^2$ region between 4 and 8~GeV$^2$.

\subsection{Special mention to $b \to c \ell \bar{\nu}$ anomalies}

\begin{figure}[t]  
\centering
\includegraphics[width=0.7\textwidth]{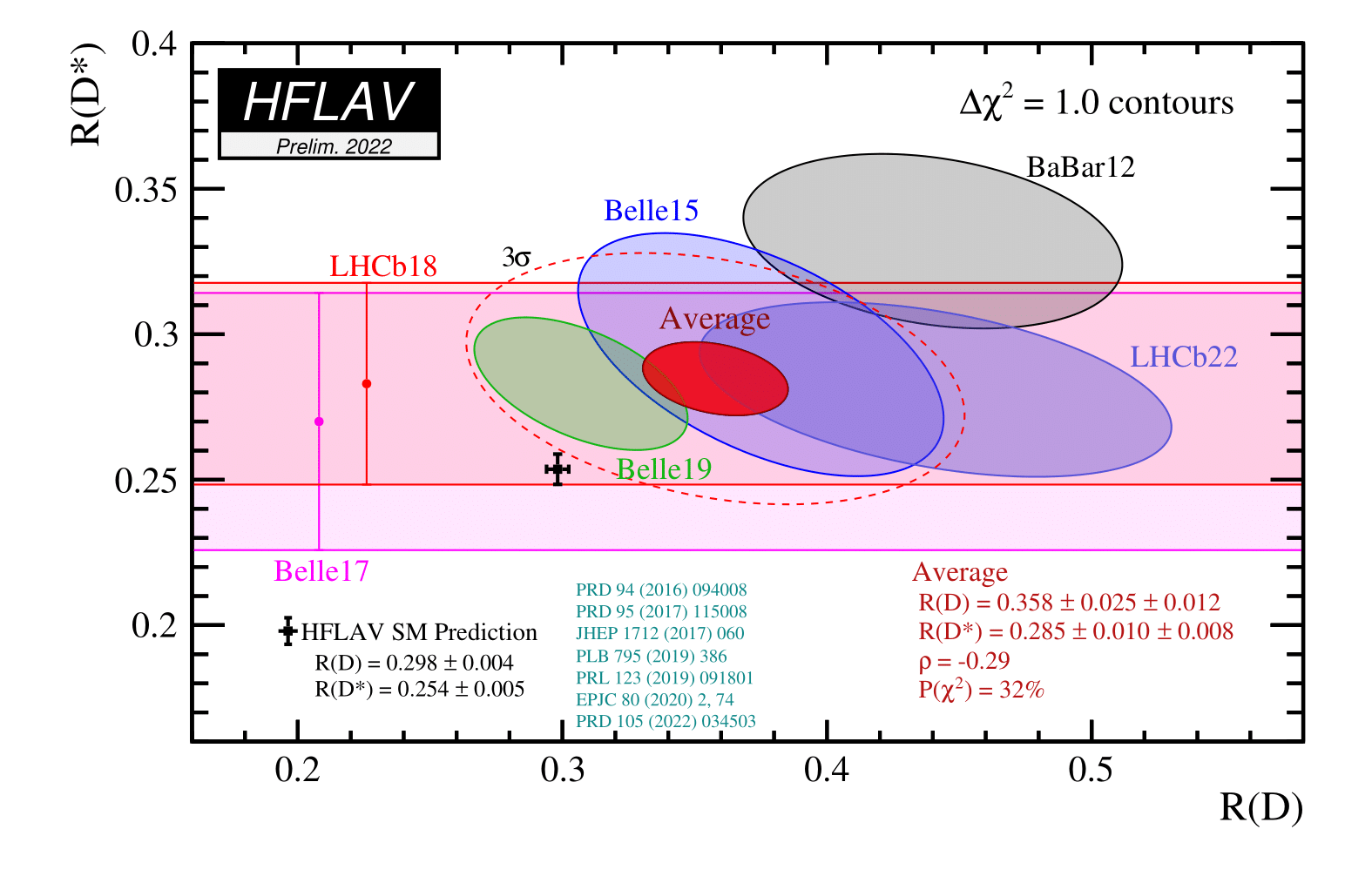}
\caption{Summary of all the published $R(D^{(*)})$ measurements~\cite{HFLAV:2022pwe}.}
\label{fig:RDRDst}
\end{figure}

A special mention among the flavour anomalies must go to a completely different class of measurements,
namely $b \to c \ell \bar{\nu}$ transitions.
These decays, ruled by tree-level processes in the Standard Model, have shown another intriguing pattern of deviations.
Similarly to Eq.~\ref{eq:RX}, one can create a Lepton Flavour Universality ratio based on the ratio 
\begin{equation}\label{eq:RD}
R_D^{(*)} = \frac{\mathcal{B}( B \to D^{(*)} \tau \bar{\nu})}{\mathcal{B}( B \to D^{(*)} \mu \bar{\nu})} \, ,
\end{equation}
which is equally well predicted by the theory and provides a fundamental test of the SM.
During the last 10 years, measurements from different experiments have shown a consistent tension with the SM prediction,
with the current world average which is found to be around three standard deviations away from the Standard Model.
Figure~\ref{fig:RDRDst} summarizes all the available measurements in the $R(D)$-$R(D^*)$ plane.
This class of measurements is particularly interesting being the only one involving $\tau$ leptons.

\section{$b \to s \ell\ell$ anomalies: Theory}

Flavour Changing Neutral governed decays have always played a leading role in the search for New Physics. In recent times the decays $B \to K \ell\ell$ and $B \to K^*(\to K\pi) \ell\ell$ have been particularly prominent. The latter 4-body decay is parametrized in terms of 4 quantities: three angles and the dilepton invariant mass squared denoted by $q^2$ (see Refs.~\cite{
Egede:2008uy,Altmannshofer:2008dz,Egede:2010zc} for definitions). Moreover, all form factors that enter the matrix elements of this decay reduce to only two quantities, called soft form factors,  in the limit that the final meson has a large 
energy~\cite{Charles:1999gy}. These quantities, of course, as any hadronic quantity, have an inherent hadronic uncertainty associated.
 Looking a bit far in prehistory, already in 1995, it was realized in Ref.~\cite{Ali:1994bf} that the zero of one observable, called forward backward asymmetry of the $B \to K^*\ell^+\ell^-$ decay exhibited an interesting property.  The observable has the peculiarity that at the value of $q^2$ determined by its zero  an exact  cancellation of soft form factors at LO occurs. This offered a strategy to identify New Physics by using a clean combination of Wilson coefficients~(see definitions below) free from soft form factors.  While 
 interesting, it is experimentally rather difficult to measure with high precision the position of this zero, besides the fact that the NLO contributions affect its position. Some years  later in Ref.~\cite{Kruger:2005ep} a change of strategy was proposed, instead of using traditional observables and looking for this zero, a method to construct a new type of observable was proposed that exhibits this property for all $q^2$. The observables constructed in this way should respect the symmetries of the distribution \cite{Egede:2010zc,Matias:2012xw}. This leads to the so-called first optimized observables~\cite{Kruger:2005ep}. Further observables were proposed in Ref.~\cite{Becirevic:2011bp} and, finally, the first complete basis was presented in Ref.~\cite{Matias:2012xw} and it was slightly modified in Refs.~\cite{Descotes-Genon:2012isb, Descotes-Genon:2013vna}. The observables of the complete basis that  parametrize the full distribution were called $P_i$\footnote{The $P$ stands for primary, like the primary colors.}.

In the previous section the experimental aspects of the different B-flavour anomalies were discussed in  detail. However, it may be interesting to add a few comments. In 2013 the first anomaly in the observable $P_5^{\prime\mu}$ was announced 
by LHCb~\cite{LHCb:2013ghj}. This is one of the most tested anomalies, not only in different updates using more and more data, but also by different experiments: Belle, ATLAS and CMS (see refs in previous section). Recently, also the charged channel has been measured by LHCb \cite{LHCb:2020gog} finding, albeit with large errors, also a discrepancy with SM. A different kind of anomaly, also discussed in the previous section, was observed in quantities called $R_{K,K^*}$ proposed in \cite{Hiller:2003js} and constructed to test violations of lepton flavour universality. These observables, given their conceptual importance, are now under strong scrutiny. Finally, the measurement of ${\cal B}_{B_s \to \mu\mu}$ by LHCb, CMS and ATLAS showed a combined deviation from SM by about 2$\sigma$. But adding the very recent measurement by CMS~\cite{CMS-PAS-BPH-21-006} (still not published) will reduce substantially the discrepancy with the SM prediction. 

 

In order to do a combined analysis of all observables governed by the $b\to s\ell\ell$ transition it is customary to introduce an effective Hamiltonian 

$$
  \mathcal H_\text{eff}^{bs\ell\ell}
=-\frac{4G_F}{\sqrt{2}} V_{tb}V_{ts}^*\sum_i {\cal C}_{i}  {\cal O}_i
  +
   \text{h.c.}
$$
where the relevant operators here are:
$$
\begin{aligned}
O^{bs}_7 &= \frac{e}{16 \pi^2} {m_b} (\bar{s} \sigma_{\mu\nu} P_{R} b) F^{\mu\nu}\,,
&
O^{\prime bs}_7 &= \frac{e}{16\pi^2} {m_b} (\bar{s} \sigma_{\mu\nu} P_{L} b) F^{\mu\nu}\,,
\\ 
O_9^{bs\ell\ell} &= \frac{e^2}{16 \pi^2}
(\bar{s} \gamma_{\mu} P_{L} b)(\bar{\ell} \gamma^\mu \ell)\,,
&
O_9^{\prime bs\ell\ell} &= \frac{e^2}{16 \pi^2} (\bar{s} \gamma_{\mu} P_{R} b)(\bar{\ell} \gamma^\mu \ell)\,,
\\
O_{10}^{bs\ell\ell} &= \frac{e^2}{16 \pi^2} (\bar{s} \gamma_{\mu} P_{L} b)( \bar{\ell} \gamma^\mu \gamma_5 \ell)\,,
&
O_{10}^{\prime bs\ell\ell} &= \frac{e^2}{16\pi^2} (\bar{s} \gamma_{\mu} P_{R} b)( \bar{\ell} \gamma^\mu \gamma_5 \ell)\,,
\\
\end{aligned}
$$
The New Physics we are searching for is encoded inside the Wilson coefficients of these operators. For this reason we split these coefficients in:
\begin{equation}
{\cal C}_i={\cal C}_i^{\rm SM}+{\cal C}_i^{\rm NP}
\end{equation}
Among them the Wilson coefficient ${\cal C}_9^{\rm eff}$\footnote{The effective superscript is used to denote that it always appear in a certain combination with the Wilson coefficients of 4-quark operators $O_{1-6}$. See Ref.~\cite{Alguero:2022wkd} for a detailed discussion on the structure of this coefficient.} plays a leading role in our understanding of the anomalies~\cite{Descotes-Genon:2013wba,Alguero:2022wkd}.

\subsection{Present: Global analyses of $b\to s\ell\ell$ transitions}

Our latest most complete and updated global analysis including 254 observables from LHCb, Belle, ATLAS and CMS was presented in Ref.~\cite{Alguero:2021anc} of which 24 are LFUV observables.
The list of observables considered includes:
\begin{itemize}
\item Optimized $P_i$ from $B^0 \to K^{*0} \mu^+\mu^-$  and $F_L$ from LHCb, ATLAS and CMS. Also the charged one $B^+ \to K^{*+} \mu^+\mu^-$ from LHCb.
\item Low-q$^2$ electronic observables coming from $B^0 \to K^{*0} e^+ e^-$.
\item Optimized $P_i$ from $B_s \to \phi\mu^+\mu^-$.
\item Branching ratios of $B_s \to \phi \mu^+\mu^-$ (LHCb), $B^{+,0} \to K^{(*)+,0}\mu^+\mu^-$ (LHCb,CMS, Belle) also inclusive ones from Babar.
\item Branching ratio of $B_s \to\mu^+\mu^-$ combined from LHCb, ATLAS and CMS (only published numbers).
\item $F_L$ and $A_{\rm FB}$ observables from CMS (neutral and charged channel).
\item LFUV observables $R_{K^{*0}}$, $R_{K^+}$, $R_{K^{*+}}$ and $R_{K_S}$ from LHCb and Belle.
 \item LFUV $Q_{4,5}=P_{4,5}^{\prime \mu}-P_{4,5}^{\prime e}$~\cite{Capdevila:2016ivx} from Belle.
 \item Radiative decays: $B \to X_S \gamma$, $B^0 \to K^{*0}\gamma$ ($A_I$ and $S_{K^*\gamma}$), $B^+\to K^{*+}\gamma$ and $B_S \to \phi \gamma$.

\end{itemize}
Of course, all available high and low-q$^2$ regions are included. Other existing analyses in the literature like the one presented in Ref.~\cite{Hurth:2021nsi} differs on  some aspects of the treatment of hadronic uncertainties but contain also all available data including the $\Lambda_b \to \Lambda\mu^+\mu^-$ decay. It is remarkable the very good agreement in terms of hierarchies of NP hypotheses and results among these two independent analyses. A subset of the data mentioned above  is used in Ref.~\cite{Altmannshofer:2021qrr} and, finally, examples of bayesian analyses are presented in Refs.~\cite{Kowalska:2019ley,Blake:2019guk} and \cite{Ciuchini:2017mik} (excluding in this latter case all high-q$^2$ data). For a more detailed comparison among the different analyses we refer the reader to Ref.~\cite{London:2021lfn}.

\subsubsection{Two sources of hadronic uncertainties for exclusive decays}

The amplitude for the decay $B \to M\ell^+\ell^-$ has the following structure in the SM: 

\begin{equation}
A(B\to M\ell\ell)=\frac{G_F \alpha}{\sqrt{2}\pi} V_{tb}V_{ts}^* [({A_\mu}+{ T_\mu}) \bar{u}_\ell \gamma^\mu v_\ell
 +{B_\mu} \bar{u}_\ell \gamma^\mu \gamma_5 v_\ell]
\end{equation}
 The study of this type of exclusive semileptonic B decays requires to control two type of hadronic uncertainties~\cite{Capdevila:2017ert}:

\begin{itemize}
\item Local contributions (more terms should be added to the amplitude if there is NP in non-SM ${\cal C}_{i}$), so called, form factors:
 \begin{eqnarray}
{A_\mu} &=& -\frac{2m_bq^\nu}{q^2} {\cal C}_7 \langle M | \bar{s}\sigma_{\mu\nu}P_R b|B\rangle
 +{\cal C}_9 \langle M | \bar{s}\gamma_\mu P_L b|B\rangle\\
{B_\mu} &=& {\cal C}_{10} \langle M | \bar{s}\gamma_\mu P_L b|B\rangle.
\end{eqnarray}
 These form factors are computed using LCSR with either B-meson or light-meson DA  plus  lattice at high-q$^2$ (or extrapolated at low-q$^2$). See Ref.~\cite{Capdevila:2017ert} for details on our treatment of form factors and correlations.
\item Non-local contributions (charm loops): 
The fact that the $T_\mu$ contribution enters the same structure as the ${\cal C}_{7,9}$ Wilson coefficients has been the source of endless discussions about its role in explaining the anomalies. However, one should not forget that this contribution is $q^2$ and helicity-dependent (in the $K^*$ case) and function of the external states of the exclusive decay. Its signature can be disentangled  with enough precision  from a universal NP contribution (same for all bins and for different exclusive decays), by comparing the values of ${\cal C}_9^{\rm NP}$ obtained in different bins. No evident signal of an extra $q^2$ contribution has been observed, besides the explicit evaluation of the long-distance contribution (one soft-gluon exchange) by Ref.~\cite{Khodjamirian:2010vf} that should be included in any analysis. Furthermore, in recent years there have been a number of different rigorous approaches~\cite{Gubernari:2020eft,Blake:2017fyh,Bobeth:2017vxj} to tackle this problem (see Ref.~\cite{London:2021lfn} for a discussion), with a common outcome, namely, the impossibility to explain the anomalies in $b\to s\mu\mu$ decays using only a SM point of view.

\begin{figure}
\hspace*{1cm}{\includegraphics[width=0.25\textwidth]{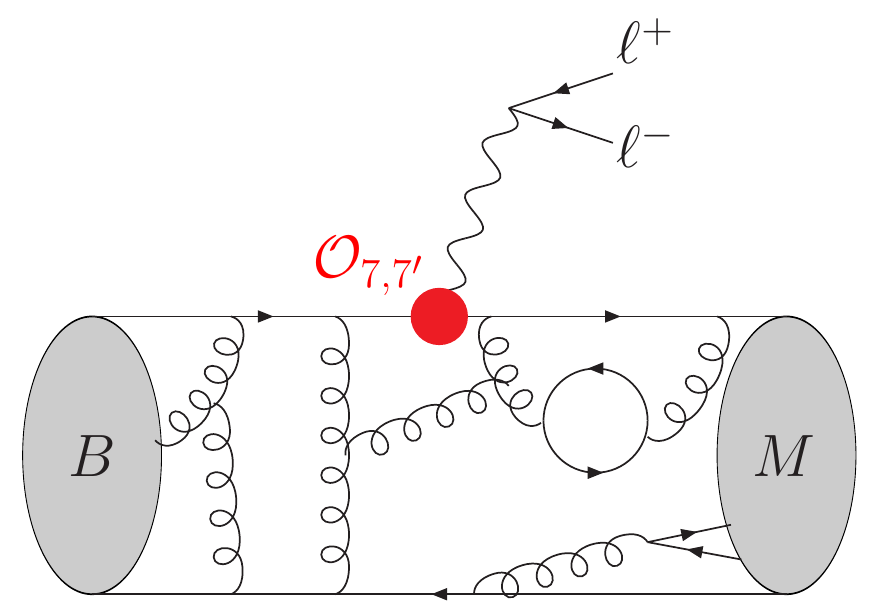}\includegraphics[width=0.25\textwidth]{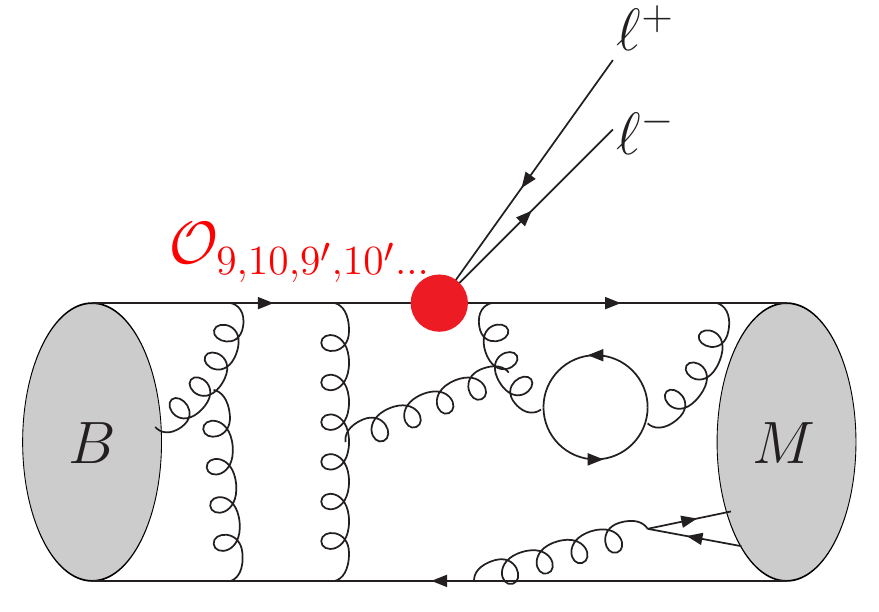}
{\includegraphics[width=0.25\textwidth]{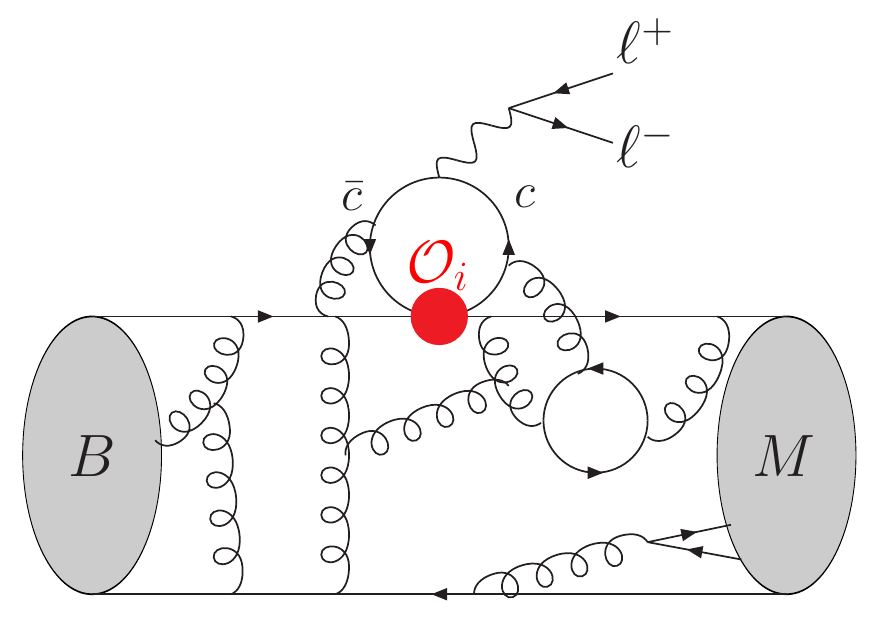}\quad\quad {}}}
{ \qquad\qquad\qquad\quad 
{\hspace*{2.7cm}\quad{Form factors (local)}}
\qquad\qquad 
\hfill{Charm loop (non-local)}} \hspace*{0.1cm} {}
\caption{Illustration of the two types of main sources of hadronic uncertainties: form factors and charm loops.}
\end{figure}    
\end{itemize}

\subsection{Results of $b\to s \ell\ell$ global fits}

The result of the global fit to one Wilson coefficient (or constrained pair) is presented in Table.\ref{tab1:table1}
and corresponds to the updated results in Ref.~\cite{Alguero:2021anc} following the methodology in Refs~\cite{Descotes-Genon:2015uva,Capdevila:2017bsm,Alguero:2019ptt}. Some remarks are in order:

\begin{table}[tbp] 
\renewcommand{\arraystretch}{1.5}
{\begin{tabular}{ccccccc} 
\hline
 & \multicolumn{3}{c}{All} &  \multicolumn{3}{c}{LFUV}\\
\hline
1D Hyp.& Best fit   &  1 $\sigma$   & Pull$_{\rm SM}$ & p-value & 1 $\sigma$  & Pull$_{\rm SM}$ \\
\hline
${\cal C}_{9\mu}^{\rm NP}$  & -1.01 &   $[-1.15,-0.87]$ &    7.0   & 24.0\,\%
 &$[-1.11,-0.65]$&  4.4   \\
${\cal C}_{9\mu}^{\rm NP}=-{\cal C}_{10\mu}^{\rm NP}$    &  -0.45 &  $[-0.52,-0.37]$ &  6.5  & 16.9\,\%
 &    $[-0.48,-0.31]$ &   5.0     \\
${\cal C}_{9\mu}^{\rm NP}=-{\cal C}_{9'\mu}$    & -0.92&   $[-1.07,-0.75]$ & 5.7  & 8.2\,\%
 &    $[-2.10,-0.98]$  & 3.2    \\
 \hline
\end{tabular}}
\caption{Most prominent 1D patterns of NP in $b\to s \mu\mu$ for the fit to All observables or only the LFUV subset.} \label{tab1:table1}
\end{table}

\begin{itemize}
        \item The $p$-value of SM hypothesis  is now {0.44\%} (2022)  for the fit ``All"
   and {0.91\%} (2022)  for the fit ``LFUV".
    \item Still tension is observed between the ``All" and ``LFUV" fit due to  their opposite preferences between the two main 1D scenarios.      A solution to this tension is discussed in section~\ref{sec:3.3}.

\end{itemize}
%

\medskip



The results for the 2D (with and without LFU) and 6D New Physics scenarios can be found in Ref.~\cite{Alguero:2021anc}. The most relevant LFU case is discussed in the next section. Finally, it is interesting to point out what the most relevant observables tell us about specific Wilson coefficients:

\begin{itemize}

\item[1]  $ {\cal B}_{B_s \to \mu^+\mu^-}$ exhibits a small (but persistent) deviation from the SM. This observable drives the contribution to  ${\cal C}_{10\mu}^{\rm NP}$ (positive and small) or ${\cal C}_{10\prime\mu}^{\rm NP}$ (negative) or a combination of both or a scalar contribution.


\item[2] $P_5^{\prime\mu}$ requires a { large} (in absolute value) negative contribution to  ${\cal C}_{9\mu}^{\rm NP}$.


\item[3] $R_X$ (with $X=K,K^*$,...) signals the { presence of LFUV} but they 
admit many solutions with $C_{9\mu}$ and $C_{10\mu}$ together with their chiral counterparts that give similar results. For this reason it is quite difficult using only these observables to disentangle the right scenario among the different possibilities.

\end{itemize}


\subsection{{Solution: LFU New Physics}}\label{sec:3.3}

\begin{figure}
\centering
\includegraphics[width=0.35\textwidth,height=3.8cm]{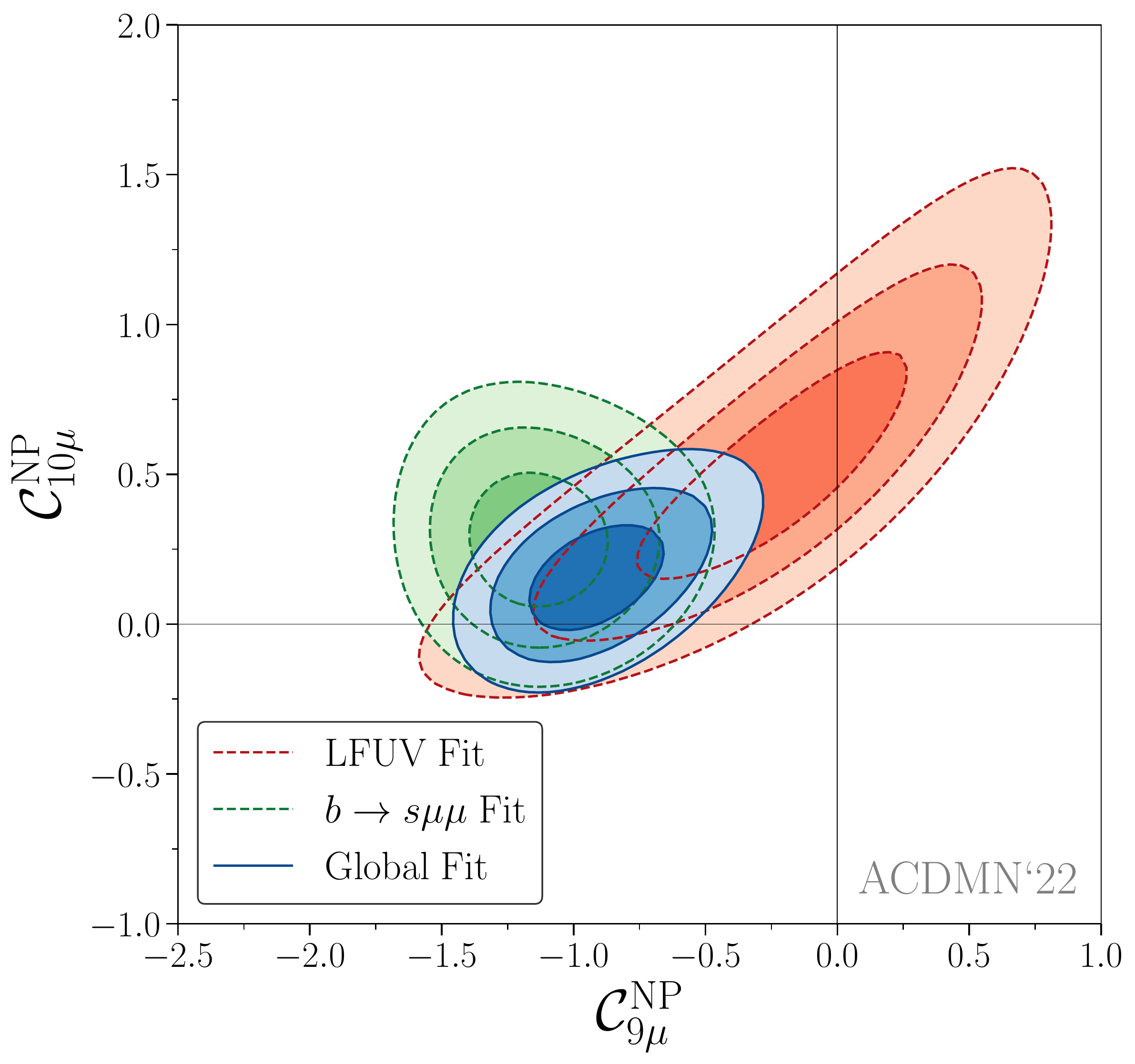}
\includegraphics[width=0.35\textwidth,height=3.8cm]{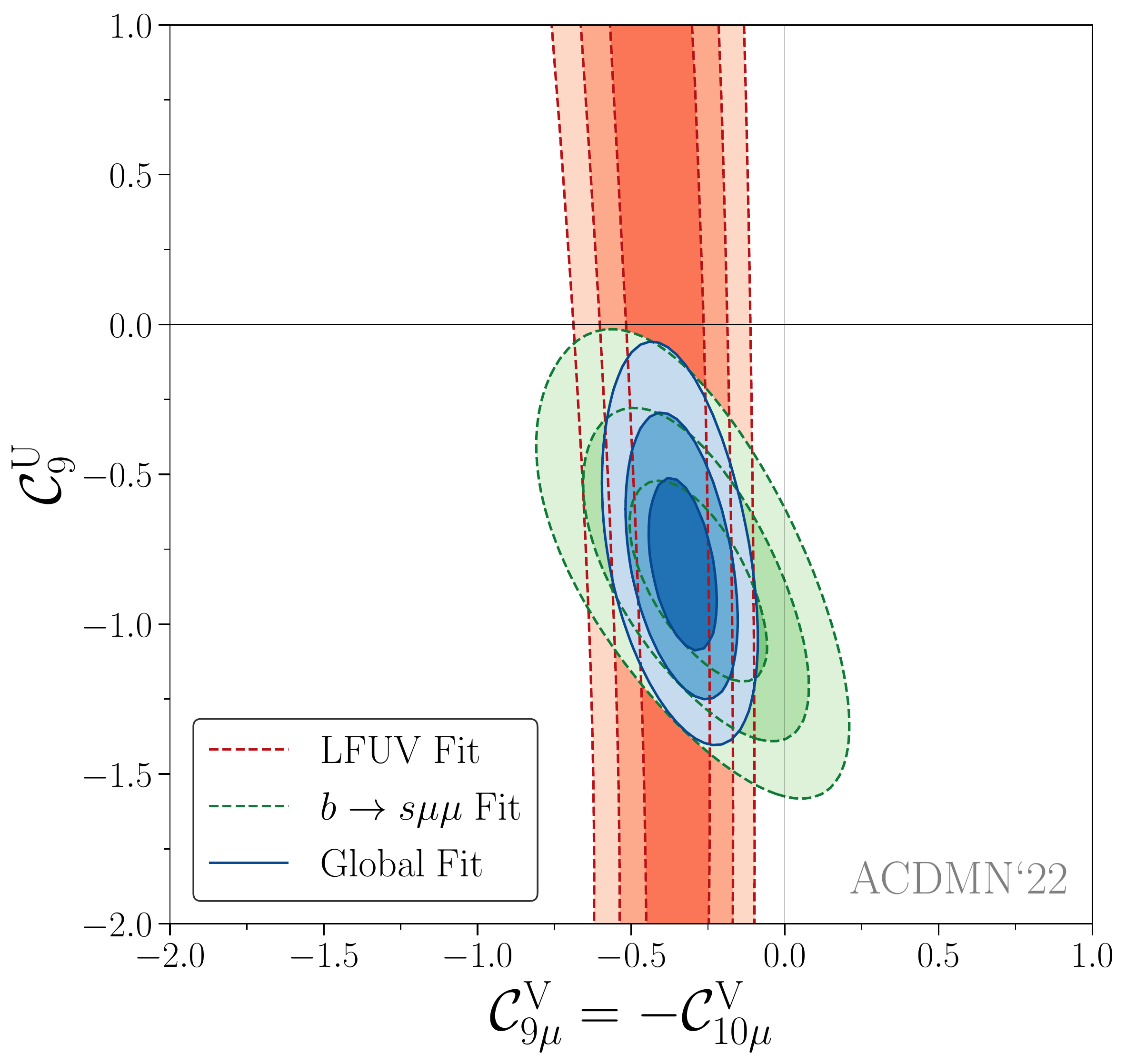}
\caption{Allowed regions in the $({\cal C}_{9\mu}^{\rm NP},{\cal C}_{10\mu}^{\rm NP})$  and $({\cal C}_{9\mu}^V=-{\cal C}_{10\mu}^V, {\cal C}_9^U)$ plane for the corresponding 2D hypotheses separating the $b\to s \mu\mu$ fit (green), the ``LFUV" fit  (red) and the ``All" fit (blue). }
\label{fig:allowed}\end{figure}

In order to solve the puzzling problem of the preference for ${\cal C}_{9\mu}$ New Physics hypothesis in ${\cal C}_{9\mu}=-{\cal C}_{10\mu}$ of the ``All fit" and the opposite situation for the ``LFUV" fit, it was proposed in Ref.~\cite{Alguero:2018nvb} to remove the hypothesis that New Physics is purely LFUV. This is implemented by splitting the Wilson coefficients in the following form
\begin{equation}
C^{\rm NP}_{ie}=C^{\rm U}_{i} \quad C^{\rm NP}_{i\mu}=C^{\rm V}_{i\mu}+C^{\rm U}_{i}
\end{equation}
where $i=9,10$. In this way, 
\begin{itemize}
\item A common New Physics contribution $C^{\rm U}_i$ to the charged leptons is introduced.
\item It allows to accommodate that LFUV New Physics prefers an SU(2)$_L$ structure while LFU New Physics is vectorial.
\end{itemize}
Fig.~\ref{fig:allowed}  illustrates that in the correlated ${\cal C}_{9\mu}-{\cal C}_{10\mu}$ scenario the fit to only $b \to s\mu\mu$ observables and the fit to only LFUV observables are in slight tension. On the contrary, in the Scenario $[{\cal C}_{9\mu}^V=-{\cal C}_{10\mu}^V, {\cal C}_9^U]$ called Scenario 8 in Ref.~\cite{Alguero:2018nvb} (or Scenario-U, due to the presence of Universal New Physics)   there is a perfect agreement between both fits, solving the above mentioned problem.

\subsection{ SMEFT connection between $b\to s\mu\mu$ \& $b\to c \ell\nu$ in Scenario-U}\label{sec:34}

Still in the New Physics hypothesis called Scenario-U it was shown in Ref.~\cite{Capdevila:2017iqn} that it is possible to establish a connection between charged and neutral anomalies using SMEFT,
%
%
%
%
%
%
%
\begin{figure}[b] \centering
{\includegraphics[width=0.3\textwidth,height=2cm]{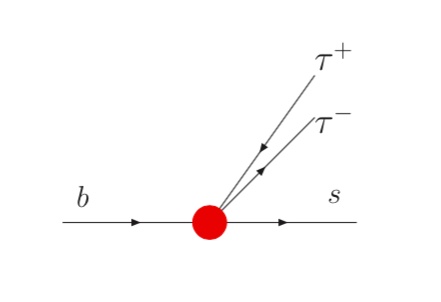}}
{\includegraphics[width=0.3\textwidth,height=2cm]{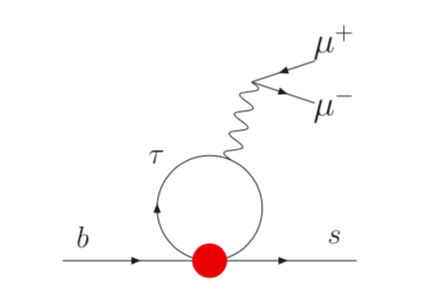}}
\caption{(Left) effective ${\cal O}_9^{\tau\tau}$ operator.
(Right) Diagram inducing a lepton universal contribution to ${\cal C}_9^U$ through a $\tau$ loop.
}\label{fig3}
\end{figure}

 $${\mathcal L}_{SMEFT}={\mathcal L}_{SM}+{\mathcal L}_{d>4}$$
 Assuming a common New Physics explanation for $R_D$, $R_{D^*}$ and $R_{J/\Psi}$ based on the SM operator $[\bar{c} \gamma_\mu P_L b][\bar{\tau} \gamma^{\mu} P_L \nu_{\tau}]$  with a similar size of the ratios to its SM predictions amounts to changing the normalization of the Fermi constant for $b \to c \tau^- \bar{\nu}_\tau$ transitions. Moreover, if this contribution is generated at a scale larger than the electroweak scale then one would need to consider only the following two operators with left-handed doublets:
\begin{equation}
{\cal O}^{(1)}_{ijkl}=[\bar{Q}_{i}\gamma_\mu Q_j][\bar{L}_k\gamma^\mu L_l]
\quad
{\cal O}^{(3)}_{ijkl}=[\bar{Q}_{i}\gamma_\mu\vec\sigma Q_j][\bar{L}_k\gamma^\mu\vec\sigma L_l]
\end{equation}

In this case the FCCC part of ${\cal O}^{(3)}_{2333}$ would be responsible for generating  $R_{D^{(*)}}$, while the FCNC part of ${\cal O}^{(1,3)}_{2333}$ with $C^{(1)}_{2333}=C^{(3)}_{2333}$ (see Ref.~\cite{Capdevila:2017iqn}) would take care of the neutral anomalies. The latter constraint among the Wilson coefficients, necessary to avoid the stringent bounds from $B\to K^{(*)}\nu\bar{\nu}$, has two important implications:
\begin{itemize}
\item It induces a huge enhancement of $b\to s\tau^+\tau^-$ governed decays.
\item Through radiative effects via a tau-loop (thanks to the insertion of the  $bs\tau^+\tau^-$ operator) it is naturally generated a LFU New Physics contribution in  ${\cal C}_{9}^{\rm U}$ (see Fig.\ref{fig3}).
\end{itemize}


\begin{figure} \centering 
{\includegraphics[width=0.4\textwidth,height=5.0cm]{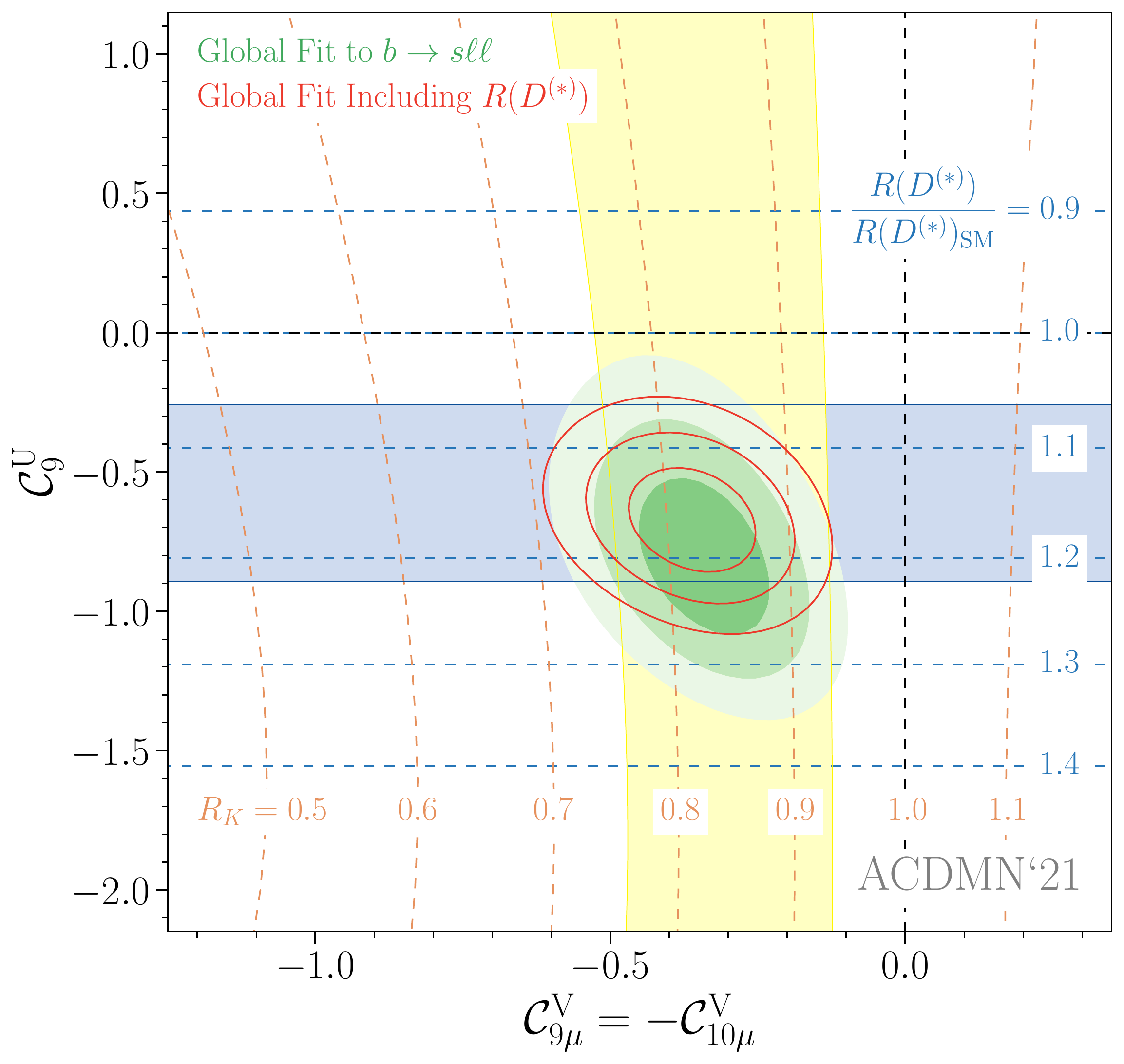}}
{\includegraphics[width=0.43\textwidth,height=5.0cm]{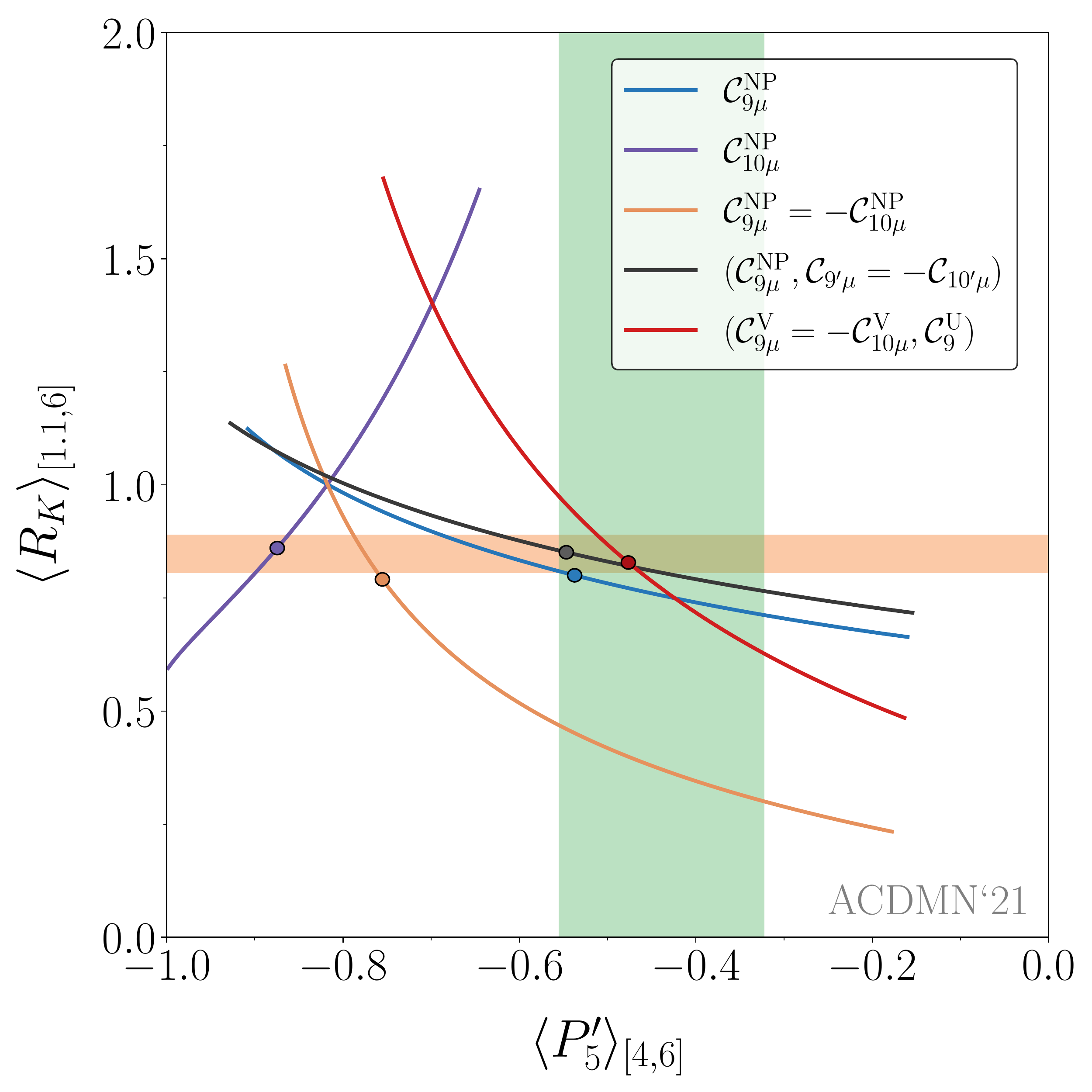}}
\caption{(Left) Preferred regions at the 1,2 and 3$\sigma$ level (green) in the 
${\cal C}_{9\mu}^V=-{\cal C}_{10\mu}^V, {\cal C}_9^U$ plane from $b \to s \ell\ell$ data. (Right) $\langle R_K \rangle^{[1.1,6]}$ versus $\langle P_5^{\prime\mu} \rangle_{[4,6]}$ in different NP scenarios.
 } \label{fig:consistency}
\end{figure}


%

\subsection{ Preferred  scenarios and consistency with: $\langle P_5^\prime\rangle_{[4,6]}$ vs $\langle R_K\rangle_{[1.1,6]}$} \label{preferred}

The New Physics hypotheses with the highest Pull$_{\rm SM}$, that provide a better explanation of the data are given in Table~\ref{tab:Tab2}. The so-called Scenario-R corresponds to the case of having a small right-handed current contribution in the Wilson coefficients ${\cal C}_{9^\prime\mu}$ and ${\cal C}_{10^\prime\mu}$ that counterbalances the effect of a large (in absolute value) vectorial ${\cal C}_{9\mu}^{\rm NP}$. The other option, most promising, because as explained above is able to connect charged and neutral anomalies, is the Scenario-U.

It is an interesting exercise to observe the consistency  of the different scenarios to explain or not the largest anomalies. Of course, this is the present picture, but it is interesting also to use it in the near future  to check if relevant changes in data can be easily accommodated or not. In Fig.~\ref{fig:consistency} it is shown that 
scenario U, R and a vectorial ${\cal C}_{9\mu}$ can explain naturally two of the largest anomalies, and scenarios like ${\cal C}_{9\mu}=-{\cal C}_{10\mu}$ or only ${\cal C}_{10\mu}$ cannot. A shift of $R_{K}$ up or down   can be naturally accommodated by the Scenario-U.

Now the last question is how to disentangle these two radically different hypotheses (Scenario-U and Scenario-R)  in the near future. This is addressed in the following section.

\begin{table}[tbp]
\label{table_scenarios}\begin{center}
{
\renewcommand{\arraystretch}{1}
\centering
\begin{tabular}{cccc} \scriptsize 
 \\
\hline
 2D Hyp.  & Best fit  & Pull$_{\rm SM}$ & p-value \\
\hline\hline 
 Scn-R $({\cal C}_{9\mu}^{\rm NP} , {\cal C}_{9^\prime\mu}=-{\cal C}_{10^\prime\mu})$ 
& (-1.15,+0.17) & 7.1 & 31.1\,\% 
\\ 
 Scn-U $({\cal C}_{9\mu}^{\rm V}=- {\cal C}_{10\mu}^{\rm V}, {\cal C}_{9}^U)$
& (-0.34,-0.82) & 7.2 & 34.5\,\%  \\
\hline
\end{tabular}
}\end{center} 
\caption{Most prominent 2D fits with highest Pull$_{\rm SM}$.}
\label{tab:Tab2}
\end{table}

\begin{figure} \centering
\includegraphics[width=7.5cm,height=4.5cm]{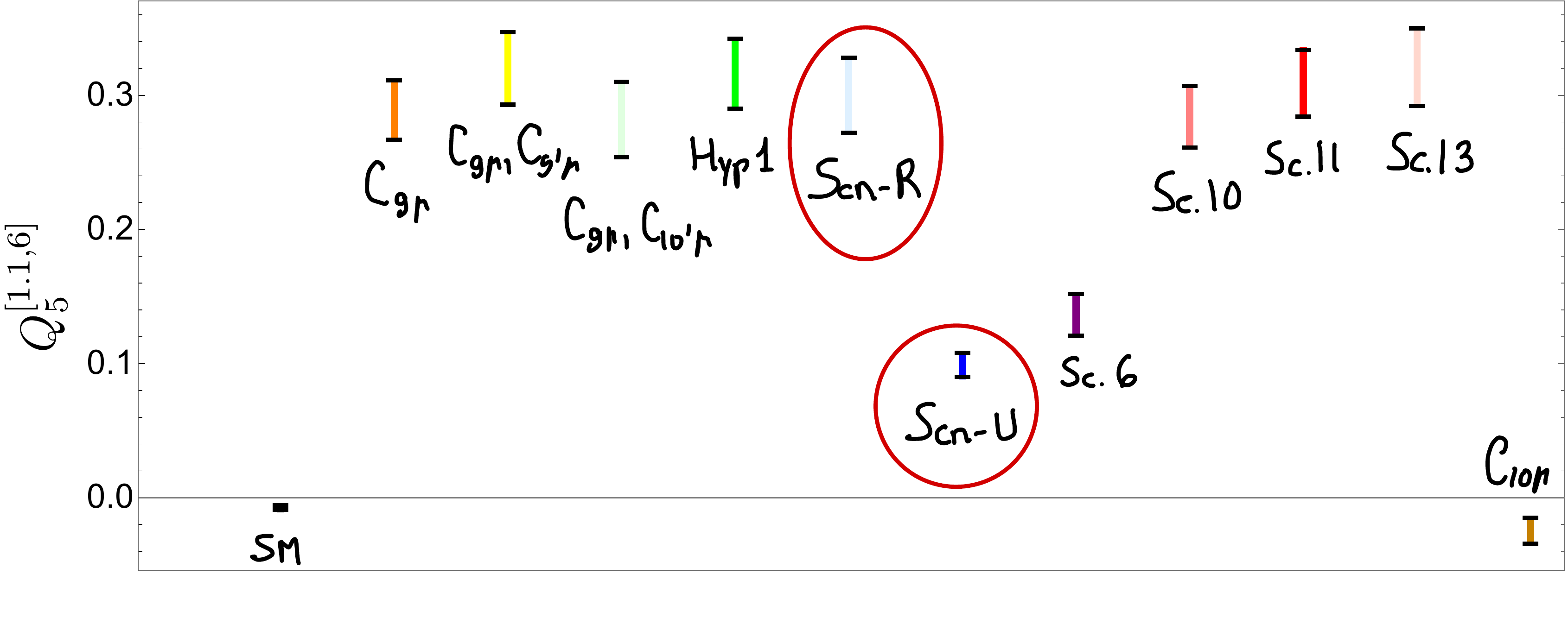}
\caption{Prediction for $Q_5^{[1.1,6]}$ in different scenarios of New Physics defined in Refs.\cite{Alguero:2021anc,Alguero:2022wkd}.}
\label{fig:Q5}\end{figure}

\subsection{Future: ${\cal C}_{9\mu}^{\rm NP}$ is the leading New Physics coefficient}
Given the dominance of the Wilson coefficient ${\cal C}_{9\mu}$ to explain the observed anomalies\footnote{After the recent measurement of $B_s \to \mu^+\mu^-$ by CMS the space for New Physics in the Wilson coefficient ${\cal C}_{10\mu}$ has been reduced a bit more.}, it is important to implement a specific analysis \cite{Alguero:2022wkd} to measure the different New Physics pieces entering this coefficient, namely,
\begin{equation} \label{eq2}
{\cal C}_{9\mu  }^{{\rm eff} } \to {\cal C}_{9\mu \, j  }^{{\rm eff} } ={\cal C}_{9\mu \, \rm pert}^{\rm SM}+ { {\cal C}_{9\mu}^{\rm NP} } + {\cal C}_{9\mu \, j}^{c\bar{c} \, B \to K^*}  \qquad  { {\cal C}_{9\mu}^{\rm NP} } =   {\cal C}_{9\mu}^V + {\cal C}_9^U 
\end{equation} 
where we address the reader to Ref.~\cite{Alguero:2022wkd} for the discussion on the details of this equation. 

In order to separate the LFUV from the LFU part one important experimental input is required 
(urgently):
 The measurement of an LFUV ${\cal C}_{9\mu}^V$ dominated observable: $Q_5=P_5^{\prime \mu}-P_5^{\prime e}$ \cite{Capdevila:2016ivx}. Indeed, this observable will not only determine ${\cal C}_{9\mu}^V$, but as it is shown in Fig.\ref{fig:Q5}, a precise measurement of $Q_5$ would be enough (and possible the only way) to disentangle the two most relevant New Physics hypotheses in Table~\ref{tab:Tab2}.


\begin{figure}
\centering
    \includegraphics[width=0.46\textwidth,height=0.23\textheight]{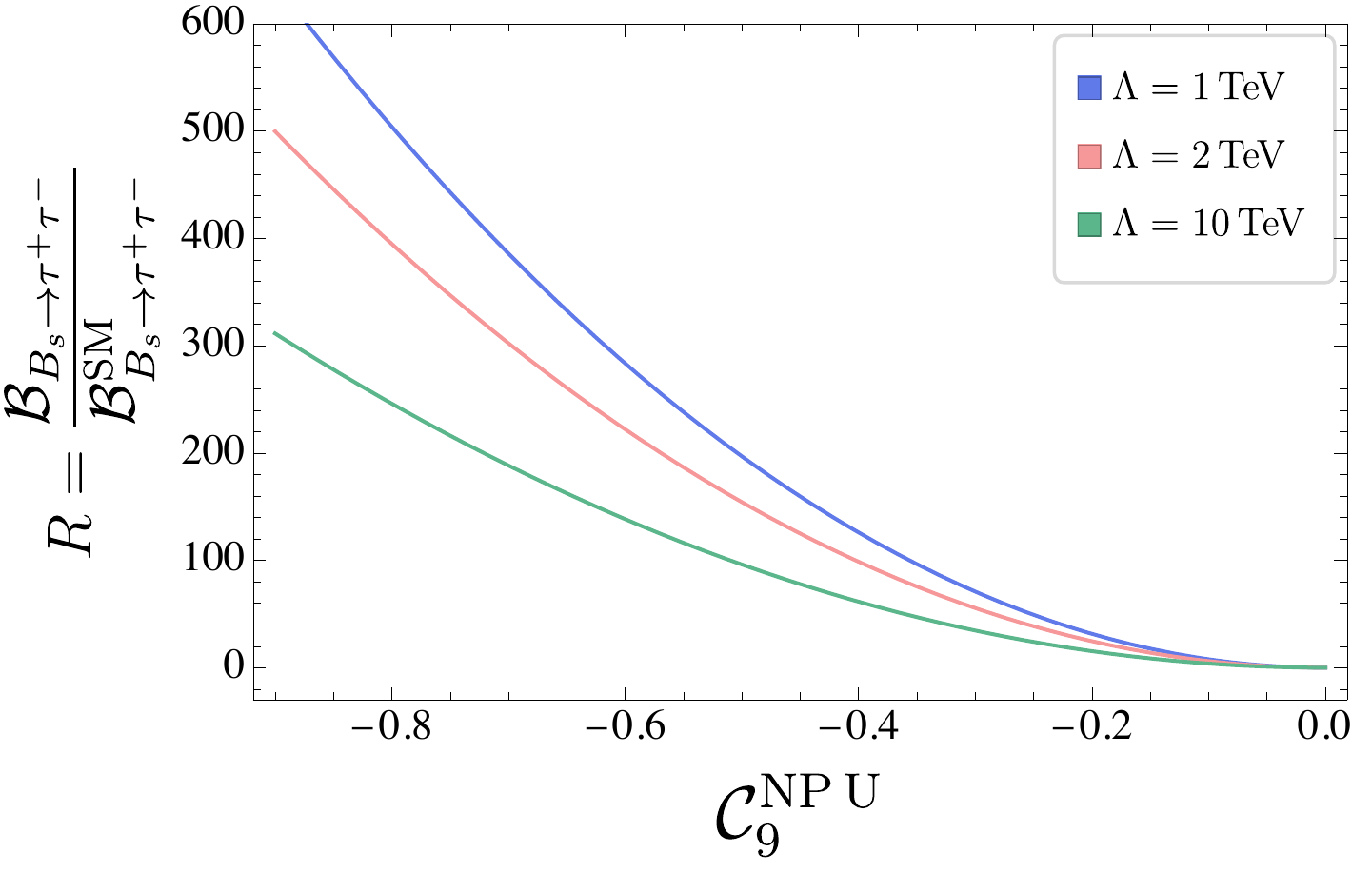}~~~~
    \includegraphics[width=0.435\textwidth,height=0.23\textheight]{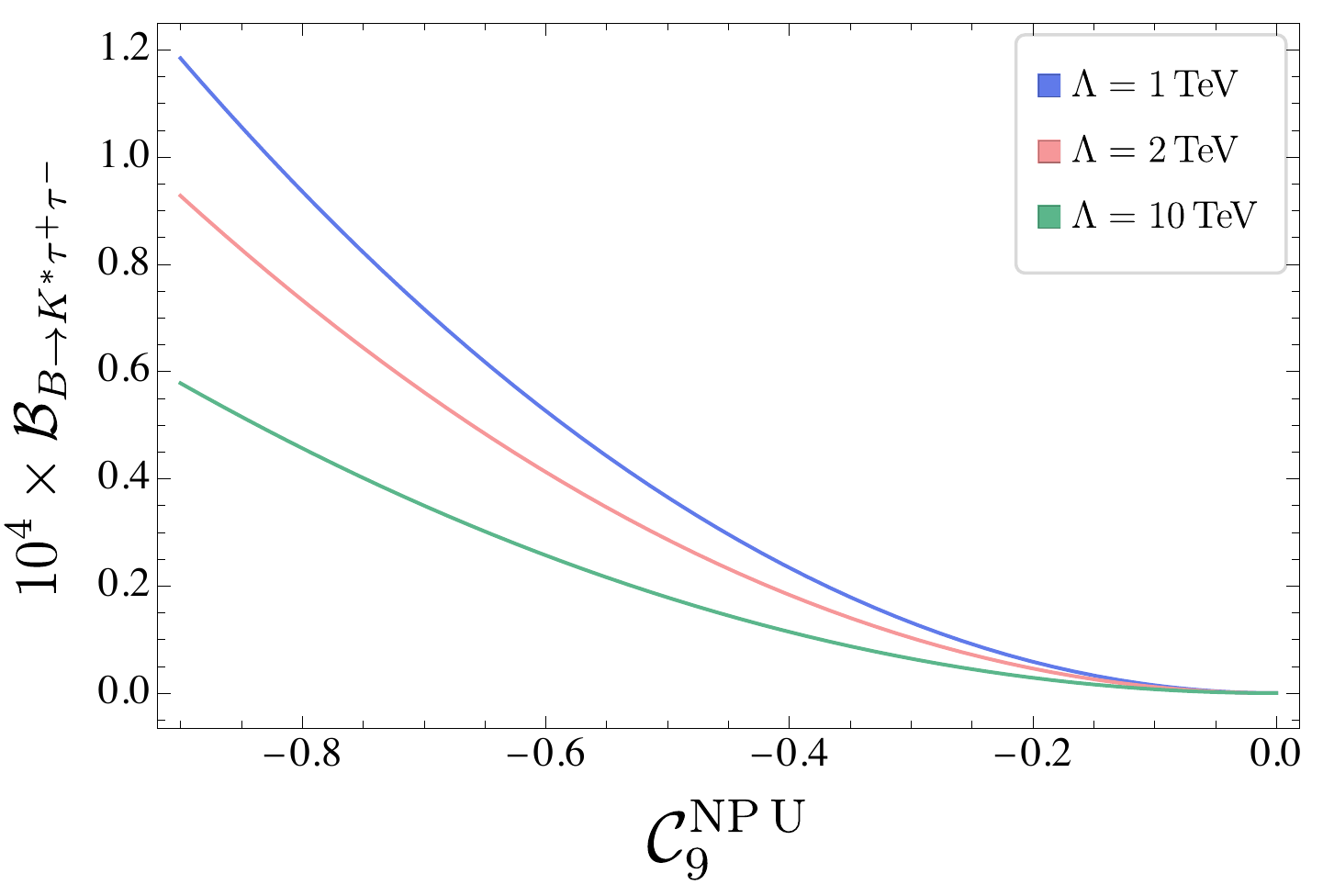}
\caption{(Left) Branching ratio of $B_s \to \tau\tau$ normalized to its SM value as a function of 
 ${\cal C}_9^{\rm NP U}$ for different values of the New Physics scale $\Lambda$. (Right) Branching ratio of $B \to K^*\tau\tau$ as a function of 
 ${\cal C}_9^{\rm NP U}$. 
 }
\label{fig:C9U}
\end{figure}
\begin{figure} \centering
\includegraphics[width=7cm]{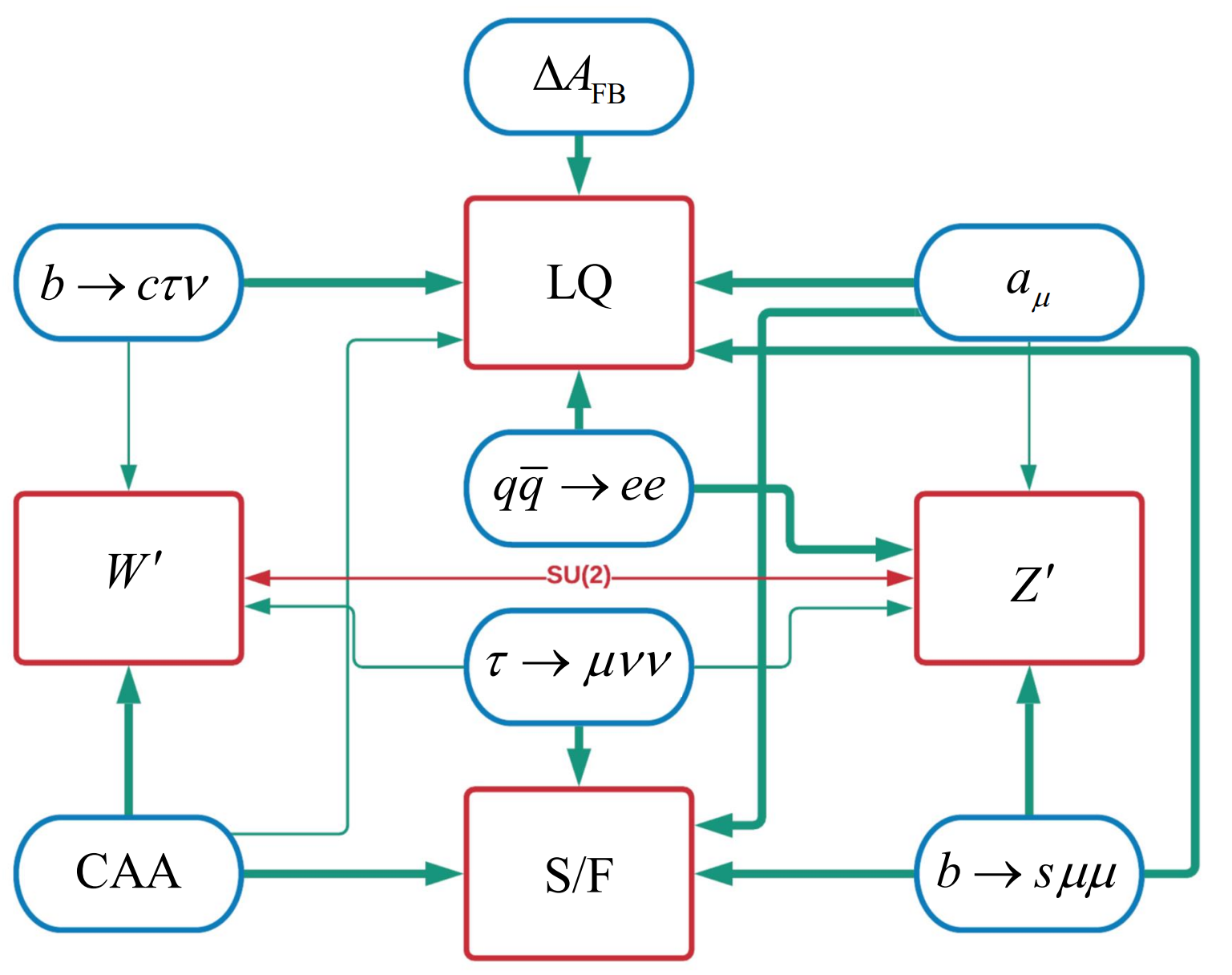}
\caption{Synthesis of possible explanations (red boxes) of the anomalies (blue boxes). The arrows indicate to which extensions of the SM the anomalies point. Thick arrows stand for probable explanations without significant experimental or theoretical shortcomings while the thin ones indicate that the new particles can only partially explain the measurement or generate problems in other observables. The red arrow indicates that the $Z^\prime$ and $W^\prime$ could be components of a single SU(2)$_L$ triplet. Figure taken from Ref.~\cite{Crivellin:2022qcj}}
\label{fig:sol}
\end{figure}
On the other side, the experimental determination of the universal piece ${\cal C}_9^{\rm U}$ comes from the measurement of $P_5^{\prime \mu}$ combined with $Q_5$ (or directly from a measurement of $P_5^{\prime e}$). It is particularly interesting exploring the possible origins of this universal piece~\cite{Alguero:2022wkd}. Among the different possibilities  the most appealing is the possibility to induce $C_9^{\rm U}$ 
via an off-shell photon penguin with the 4-fermion operator ${\bar s}\gamma^\mu P_L b \bar{f} \gamma^\mu f$ being f=$\tau$ as mentioned in Section~\ref{sec:34}. Therefore this possibility establishes a link between ${\cal C}_9^{\tau\tau}$ and ${\cal C}_9^{\rm U}$. 
Moreover, if assumed that leptoquarks are the most plausible solution given that 
they are the only particles able to generate a large ${\cal C}_9^{\tau\tau}$ without being in conflict with other processes, we will naturally have also ${\cal C}_{10}^{\tau\tau}$ with two possibilities:

\begin{itemize}

\item ${\cal C}_{9}^{\tau\tau}={\cal C}_{10}^{\tau\tau}$ in case of $S_2$ LQ

\item ${\cal C}_{9}^{\tau\tau}=-{\cal C}_{10}^{\tau\tau}$ in case of $U_1$ or $S_1+S_3$ LQs

\end{itemize}
Then automatically a measurement of ${\cal B}_{B_s \to \tau^+\tau^-}$ and/or ${\cal B}_{B \to K^{(*)}\tau^+\tau^-}$ will give us information on the New Physics entering ${\cal C}_{9}^{\rm U}$ as shown in Fig.~\ref{fig:C9U}.

\subsection{Possible particle solutions}

Finally, if one considers a larger set of anomalies (see, for instance, Ref.~\cite{Crivellin:2020oup}) and try to establish links among them in terms of new particles with their corresponding new interactions one arrives to a diagram as shown in Fig.\ref{fig:sol} (see Ref.~\cite{Crivellin:2022qcj} for details).


\section{$g-2$ anomaly: lattice}
It is well known \cite{Dirac:1928ej} that fundamental non-interacting Dirac particles have a gyromagnetic factor of $g=2$. The deviation from this value due to interactions \cite{Schwinger:1948iu} is quantified by $a=(g-2)/2$. Of particular concern is the case of the muon and the corresponding $a_\mu$.

Experimentally, the value of $a_\mu$ is known to an excellent precision. The most recent determination \cite{Muong-2:2021ojo} agrees very well with \cite{Muong-2:2006rrc} and the current combined experimental value for $a_\mu$ is \cite{Muong-2:2021ojo}
\begin{equation}
 a_\mu^\text{exp} = 16 592 061(41) \times 10^{-11}.
\end{equation}
This corresponds to a deviation of $0.35 \ \mathrm{ppm}$ from the free value.

\begin{figure}
 \centering
 \includegraphics[width=0.4\textwidth]{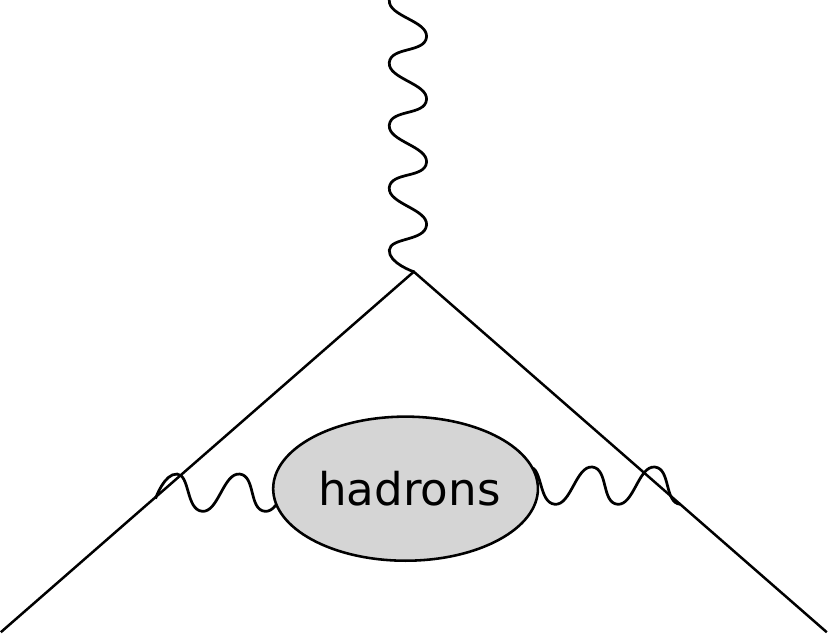}
 \caption{\label{fig_hvp_diag}
 The diagram corresponds to the leading order hadronic vacuum polarization (HVP) diagram contributing to $a_\mu$. The straight line correspond to muons and the curvy lines to photons.
 The grey blob corresponds to non-perturbative hadronic physics.
 }
\end{figure}

In the theoretical determination of $a_\mu$, many different contributions enter. Currently, a large part of the uncertainty comes from the determination of the hadronic vacuum polarization (HVP) contribution, but also the hadronic light by light scattering (HLbL) is relevant. In addition, electroweak and QED parts enter but with much smaller uncertainty. Here, we focus on the HVP determination. The leading order HVP diagram is shown in figure \ref{fig_hvp_diag}. The theoretical evaluation of this diagram is complicated by the fact that the hadronic physics contained in it is non-perturbative.

Many determinations of the HVP contribution rely on dispersive techniques. They use experimental knowledge of the hadronic $R$-ratio, the ratio of the cross section of $e^+e^- \rightarrow \text{hadrons}$ and  $e^+e^- \rightarrow \mu^+\mu^-$, to determine $a_\mu$. In \cite{Aoyama:2020ynm}, the status of these calculations and other possible approaches have been reviewed and a theoretical value of
\begin{equation}
  a_\mu^\text{LO-HVP,whitepaper} = 693.1(4.0) \times 10^{-10}
  \label{eq_hvp_whitepaper}
\end{equation}
is calculated from dispersive calculations \cite{Davier:2019can,Keshavarzi:2019abf,Colangelo:2018mtw,Hoferichter:2019mqg}. Together with estimates of the other contribution, which are also reviewed in \cite{Aoyama:2020ynm}, this yields the result
\begin{equation}
 a_\mu^\text{whitepaper} = 116 591 810(43) \times 10^{-11}
\end{equation}

Recently, lattice computations have become competitive in the determination of the HVP contribution. Conceptually, since it allows to solve the theory of the strong interaction non-perturbatively, lattice QCD+QED is very well suited to determine the HVP contribution.
The challenge lies in the required precision; to be competitive with the dispersive calculation, a subpercent precision is required. Consequently, all sources of uncertainty, both statistical and systematic, must be controlled to sufficient accuracy. In \cite{Borsanyi:2020mff} a lattice calculation of the leading order HVP contribution with an subpercent error has been presented. There, the leading order HVP contribution was determined to be
\begin{equation}
 a_\mu^\text{LO-HVP,BMWc.} = 707.5(5.5) \times 10^{-10},
\end{equation}
a larger value than the average of \cite{Aoyama:2020ynm} shown in eq.~(\ref{eq_hvp_whitepaper}) based on dispersive results.
\begin{figure}
 \centering
 \includegraphics[width=0.6\textwidth]{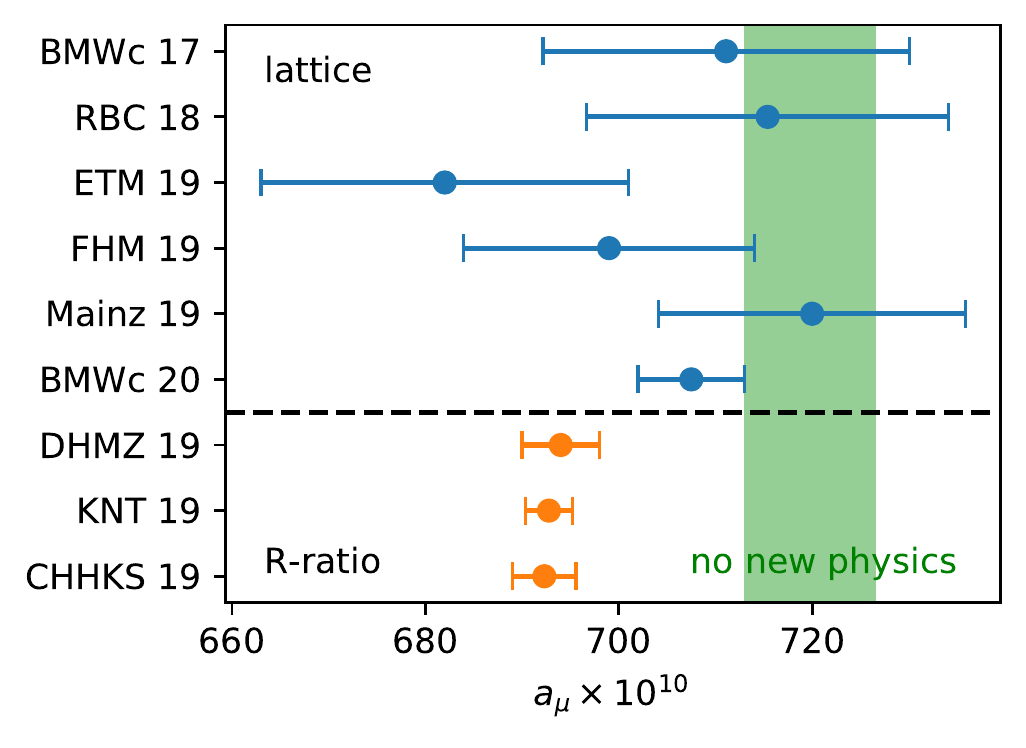}
 \caption{\label{fig_main_cmp}
 Comparison of recent lattice calculations \cite{Borsanyi:2020mff,Gerardin:2019rua,FermilabLattice:2019ugu,Giusti:2019xct,RBC:2018dos,Budapest-Marseille-Wuppertal:2017okr} and dispersive calculations \cite{Davier:2019can,Keshavarzi:2019abf,Colangelo:2018mtw,Hoferichter:2019mqg} (labeled as R-ratio) of the leading order HVP contribution to $a_\mu$.  (Source: \cite{Borsanyi:2020mff})
 }
\end{figure}
A comparison of this result with other lattice calculations \cite{Gerardin:2019rua,FermilabLattice:2019ugu,Giusti:2019xct,RBC:2018dos} and dispersive determinations \cite{Davier:2019can,Keshavarzi:2019abf,Colangelo:2018mtw,Hoferichter:2019mqg} can be found in figure \ref{fig_main_cmp}.

\begin{figure}
 \centering
 \includegraphics[width=0.7\textwidth]{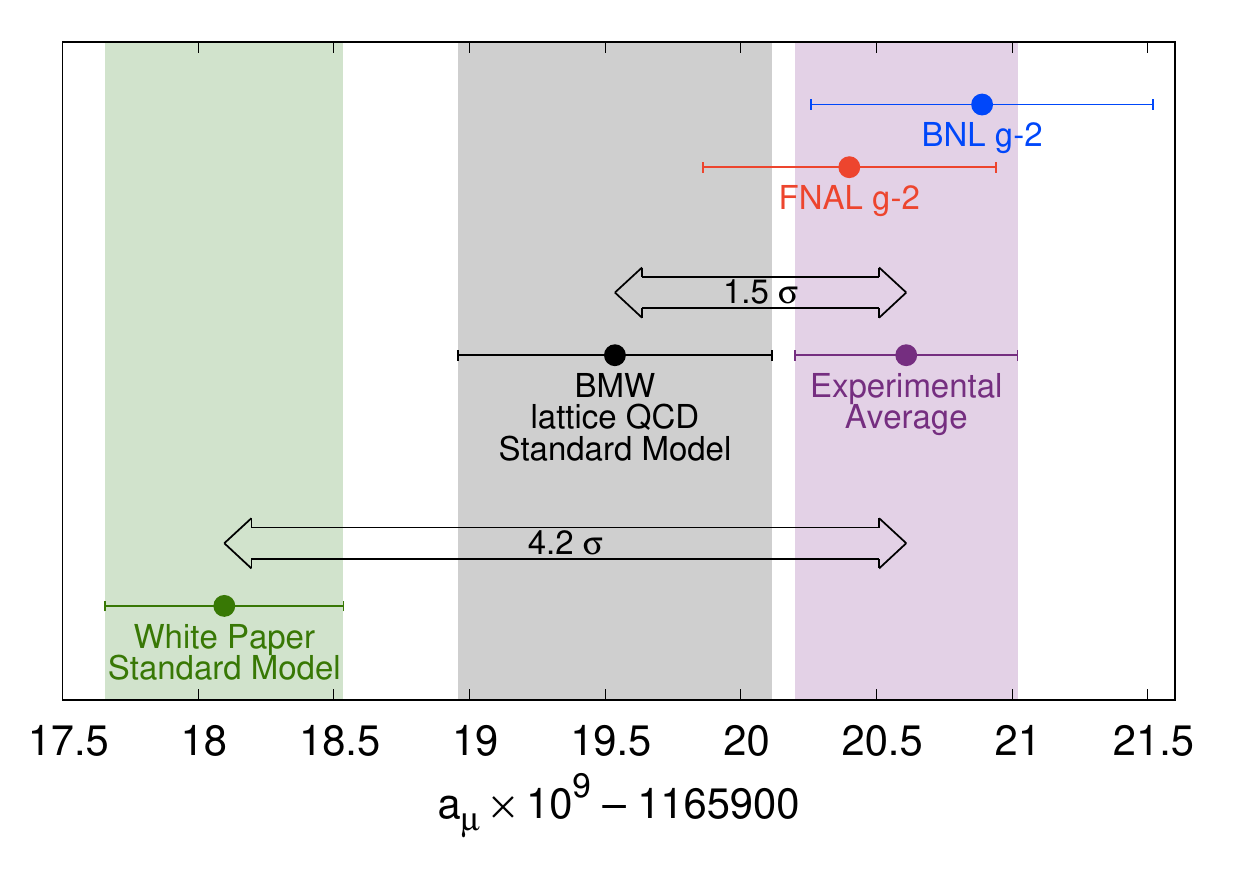}
 \caption{\label{fig_exp_vs_bmw}
 Comparison of the experimental determination of $a_\mu$ \cite{Muong-2:2006rrc,Muong-2:2021ojo} with the lattice result of the BMW collaboration \cite{Borsanyi:2020mff} and the white paper result \cite{Aoyama:2020ynm}
 }
\end{figure}
There is a tension between the experimental determinations of $a_\mu$ and the value calculated in \cite{Aoyama:2020ynm} of about $4.2 \, \sigma$. When, however, using the result of \cite{Borsanyi:2020mff} for the the leading order HVP contribution instead of the combination of dispersive calculations, this tension is only about $1.5\,\sigma$. The situation can be seen in figure \ref{fig_exp_vs_bmw}.

\subsection{The BMWc lattice calulation}
In case of the lattice computation \cite{Borsanyi:2020mff}, key elements to achieve the necessary precision where the determination of the lattice scale, the extrapolations to the infinite volume and continuum limits, the reduction of numerical noise and the inclusion of isospin breaking effects due to QED and strong isospin breaking. Here only a selection of relevant aspects can be discussed.

A large part of the calculation was carried out on a set of gauge configurations generated with Symmanzik improved action \cite{Luscher:1984xn} for the $SU(3)$ gauge fields and $2+1+1$ flavours of staggered fermions based on 4 times stout smeared \cite{Morningstar:2003gk} gauge fields. For the reference volume with a spatial extend of $L\sim 6 \, \mathrm{fm}$, 27 configurations where generated that closely scatter around the physical point. The set of ensembles features 6 lattice spacings ranging from $0.12 \, \mathrm{fm}$ to $0.06 \, \mathrm{fm}$. On top of these configurations, $U(1)$ gauge fields in the $\mathrm{QED_L}$ scheme \cite{Hayakawa:2008an} where generated. Further configurations with a different action where used for the finite volume corrections.

For the calculation of $a_\mu$, the time momentum representation \cite{Bernecker:2011gh} was used in which the magnetic moment can be written as a integral
\begin{equation}
 a_\mu = \alpha^2 \int_0^\infty \mathrm dt \, K(t) G(t) 
 \label{eq_a_mu_int_lattice}
\end{equation}
where $G(t)$ is the euclidean twopoint correlation function
\begin{equation}
 J_\mu = \frac{1}{3e}\int \mathrm d^3x \, \langle J_\mu(\vec x,t) J_\mu(\vec 0,0) \rangle.
\end{equation}
Here, $J_\mu$ is the quark electromagnetic current and $K(t)$ is a known weight function.
As it is customary in lattice computations, $a_\mu$ is split into light quark ($a_\mu^\text{light}$), disconnected ($a_\mu^\text{disc}$), strange ($a_\mu^\text{disc}$) and ($a_\mu^\text{charm})$ quark parts which can be computed separately. 

For the scale setting, the mass of the omega baryon was measured on the lattice and compared with the experimental value \cite{ParticleDataGroup:2018ovx} to determine the lattice spacing. A careful analysis of the relevant correlation function allowed to determine the scale with a precision of a few permil and to calculate the intermediate scale $w_0$ \cite{Borsanyi:2021zs} with a uncertainty of four permil.

To extrapolate to the infinite volume limit, in addition to the reference volume of $L_\text{ref}\sim 6 \, \mathrm{fm}$ used for the majority of the calculation, a big volume with $L_\text{big}\sim 11\, \mathrm{fm}$ was introduced. The difference in $a_\mu$ between these volumes has been calculated with an action that has highly suppressed taste breaking effects. The remaining small correction was calculated using two loop staggered chiral perturbation theory.
The validity of that method has been verified by comparing its predictions for the difference between the large volume and the reference volume with the numerical lattice results. Also, other models \cite{Hansen:2019rbh,Chakraborty:2016mwy,Sakurai:1960ju,Jegerlehner:2011ti,Bernecker:2011gh} for finite volume effect have been studied as a comparison.

The staggered fermion formulation used introduces taste breaking correction which strongly affect the continuum extrapolation of the light quark contribution. Therefore, taste breaking effects where reduced by subtracting the difference between the RHO model \cite{Jegerlehner:2011ti,Chakraborty:2016mwy} and its staggered variation (SRHO) in various time ranges. Other models for taste breaking effects where also studied and the systematic uncertainty in the taste breaking correction has been estimated by the comparison to NNLO staggered chiral perturbation theory in the timerange where a correction has to be applied.

The statistical noise was reduced first by utilizing, for the inversion of the Dirac operator, explicitly determined low modes \cite{Neff:2001zr,xQCD:2010pnl} and by using a method called truncated solver method \cite{Bali:2009hu} or all mode averaging \cite{Blum:2012uh}. Also, in the case of the light quark and disconnected contribution, upper and lower bounds \cite{RBRCworkshop,Borsanyi:2016lpl} for the euclidean correlation function of the quark electromagnetic current where used.

Isospin breaking in QCD+QED where calculated by a Taylor expansion \cite{deDivitiis:2013xla} around isospin symmetric QCD. Here, the expansion was done separately for the electric charge of the see ($e_s$) and the valence quarks ($e_v$). For the final result, $e_v=e_s=e$ was used.
Derivatives of the expectation value of observables with respect to the valence charge, the see charge, and the light quark mass splitting where calculated by taking either analytical derivatives or by taking finite difference, depending on what was suitable for the term in question.

The interpolation to the physical point and the extrapolation to the continuum was achieved by employing global fits to the lattice data. In these fits, various aspects of the calculation where varied. For example in case of the light quark mass contribution, roughly 0.5 million fits where used. These fits where combined using an extended version of the histogram method \cite{Borsanyi:2014jba} to determine the central value and systematic uncertainty.

\subsection{Comparison of window quantity}
\begin{figure}
 \centering
 \includegraphics[width=0.8\textwidth]{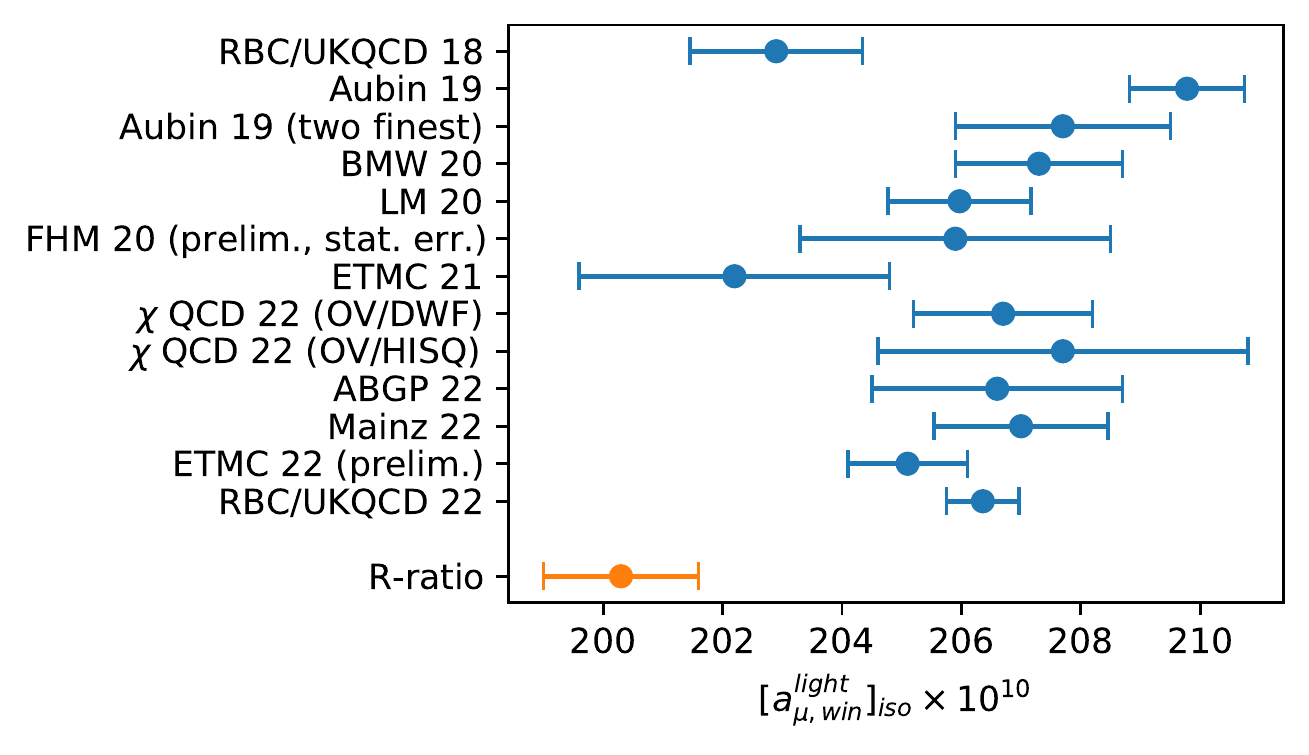}
 \caption{\label{fig_window_cmp}
 The isospin symetric, light quark component of the window quantity as determined by various lattice collaborations (\cite{RBC:2018dos} (RBC/UKQCD 18), \cite{Aubin:2019usy} (Aubin 19), \cite{Borsanyi:2020mff} (BMW 20), \cite{Lehner:2020crt} (LM 20), \cite{FHMwindow} (FHM 20), \cite{Giusti:2021dvd} (ETMC 21), \cite{Wang:2022lkq} ($\chi$QCD 22), \cite{Aubin:2022hgm} (ABGP 22), \cite{Ce:2022kxy} (Mainz 22), \cite{ETMC22window} (ETMC22), \cite{RBCUKQCD22window} (RBC/UKQCD 22)) and using the R-ratio \cite{Borsanyi:2020mff}.}
\end{figure}

To compare different lattice calculations, often a related, but simpler to calculate quantity $a_\mu^\text{win}$ \cite{RBC:2018dos} is used. It can be constructed by augmenting the weight function in eq. (\ref{eq_a_mu_int_lattice}) by the window function
\begin{equation}
 W(t,t_1,t_2) = \theta(t,t_1,\Delta)-\theta(t,t_2,\Delta)
 \ \ \ \text{with} \ \ \ 
 \theta(t,t',\Delta) = \frac{1}{2}\left(1+\tanh\left(\frac{t-t'}{\Delta}\right)\right)
\end{equation}
It is customary to use $t_1 = 0.4\,\mathrm{fm}$, $t_2=1.0\,\mathrm{fm}$, and $\Delta=0.15\,\mathrm{fm}$, although other choices are also used. This particular choice of parameters strongly suppresses the short and long distance part of the correlation function $G(t)$. These suppressed parts are heavily affected by finite lattice spacing effects in the short distance case and statistical uncertainty, taste breaking and the finite volume in the long distance case.
As a consequence, the window quantity allows to compare lattice computations while being significantly easier to calculate with high precision than $a_\mu$ itself.
Many lattice collaborations have shown results \cite{RBC:2018dos,Aubin:2019usy,Borsanyi:2020mff,Lehner:2020crt,FHMwindow,Giusti:2021dvd,Wang:2022lkq,Aubin:2022hgm,Ce:2022kxy,ETMC22window,RBCUKQCD22window} for the light quark, isospin symmetric part of this standard window. A overview can be found in figure \ref{fig_window_cmp}.

\section{$M_W$ anomaly: Experiment}

The $W$ boson plays an important role in the well-tested Standard Model theory of 
fundamental particles. Since the discovery of the weak force in radioactive decays of nuclei  
more than 100 years ago, this interaction has intrigued us and provided new insights into fundamental 
 particles and their interactions at the smallest distances. The weak nuclear force revealed 
 that nature is parity-violating and that masses of fundamental particles are emergent
 properties arising from their interaction with the 
 Higgs condensate. These properties are embodied by the $W$ boson, one of the mediators of the weak force. 

The latest, high-precision
 measurement~\cite{CDFmW2022} of the mass of the $W$ boson ($M_W$) by the CDF experiment at Fermilab 
 is in significant tension with its accurately  calculated Standard Model value. This is the latest surprise in fundamental physics and promises further revelations. This 
 measurement confirms the trend that almost all previous $M_W$ measurements have been higher than the expectation. 

The mass of the $W$ boson is theoretically linked to the $Z$ boson mass and other
  parameters.  Since $M_Z$ has been precisely measured, $M_W$ is a probe of other measurable and hypothetical parameters. 
  
  Following the discovery of the $W$ and $Z$ bosons, their masses were measured with a precision of about 1\%. The LEP and Tevatron experiments exceeded 0.1\% precision in the $W$ boson mass, and LEP achieved a precision of 23 parts per million in the mass of the $Z$ boson, over the next two decades. 
 Between 1995 and 2000, the Tevatron was 
 upgraded to deliver a factor of 100 more data than its 1992-1995 run. Using the steadily increasing data statistics between 2000 and 2011, CDF and D0 measured 
 the $W$ boson mass more precisely, reaching 0.02\% in 2012. At the LHC, the ATLAS experiment achieved a similar precision
 in 2018, and the LHCb experiment achieved 0.04\% in 2021. 
 CDF's 2022 result from its  
 complete Tevatron dataset has a precision of almost 0.01\%. 

 Most of the  complications in the $M_W$ measurement arise from the presence of the 
 neutrino in the $W$ boson decay, which invisibly carries away half of the boson's rest-mass energy. The partial inference of the neutrino momentum is one of the reasons that each iteration of the data analysis required many years. 

 The calibration of the momentum of the charged lepton 
 (electron and muon) emitted in the $W$ boson decay is a crucial aspect of the analysis. A unique feature of the CDF calibration procedure
 is the application of fundamental principles of operation of the drift chamber and the electromagnetic calorimeter. The CDF drift chamber yields the calibrated track momentum of  charged particles. CDF deployed  
 a special technique to measure the positions of the 30,240 sense wires in the drift chamber to a
 precision of $\approx 1 ~ \mu$m, using an in-situ control sample of cosmic rays. This procedure  
 eliminated many biases in track parameters. The magnetic
 field and its spatial non-uniformity is calibrated using the reconstructed invariant mass of $J/\psi \to \mu \mu$ decays. Momentum-dependent effects such as the ionization energy
 loss in the passive material are also measured precisely. The most powerful cross-checks of these calibration procedures are the demonstration
 of the consistent measurement of the $\Upsilon \to \mu \mu$ mass and of the 
 $Z$ boson mass in both the dimuon and dielectron channels. The latter also exploits a careful {\sc geant4} calculation
 of the calorimeter response and its calibration from in-situ control samples of data. Amongst all hadron-collider 
 measurements of the $W$ boson mass, the three recent CDF measurements are unique in 
 demonstrating 
 these high-precision calibrations and cross-checks from first principles. Other hadron collider measurements have solely calibrated on the reconstructed $Z$ boson mass.  The CDF experiment also demonstrates consistency between the electron- and muon-channel measurements of $M_W$ and between the $M_W$ values extracted from the distributions of the transverse mass, 
 the transverse momentum of the charged lepton and the inferred transverse momentum of the neutrino. The latter provides a cross-check of the calibration of the hadronic recoil. 
 
Precise measurements of $M_W$ are possible in the future. The LHC experiments can be precisely calibrated using in-situ
 control samples of data. Their datasets are two orders of magnitude larger than the Tevatron, and
 another order of magnitude increase is expected. A low-luminosity run to minimize pileup is being considered. An $e^+ e^-$ Higgs/electroweak factory will be the ultimate precision machine
 for $W$, $Z$ and Higgs boson properties and for following up on the $M_W$ anomaly.
\section{$M_W$ anomaly: Theory}

As we discussed in the last section, the CDF measurement of the W-boson mass is very exciting. Taken at face value, it implies significant tension with the Standard Model (SM) prediction~\cite{Haller:2018nnx} as well as with earlier, less precise, determinations~\cite{ATLAS:2017rzl,LHCb:2021bjt}. 
While the impact of potential beyond-the-SM (BSM) physics on the $W$-boson mass has previously been considered in the literature \cite{Lee:1977tib,Shrock:1980ct,Shrock:1981wq,Bjorn:2016zlr,Bryman:2019bjg,Crivellin:2021njn}, the new measurement has sparked a lot of interest to understand its BSM implications.
In particular, several groups studied the $W$-boson mass anomaly (as we will call this from now on) in terms of the Standard Model Effective Field Theory (SMEFT) under various assumptions to limit the number of independent operators and associated Wilson coefficients.

Most effort has been focused on explanations through oblique parameters \cite{Lu:2022bgw,Strumia:2022qkt,Paul:2022dds,Asadi:2022xiy} probing universal theories \cite{Kennedy:1988sn,PhysRevLett.65.964,Barbieri:2004qk}. Refs.~\cite{deBlas:2022hdk,Fan:2022yly,Bagnaschi:2022whn,Gu:2022htv,Endo:2022kiw,Balkin:2022glu} went beyond this approach by using a more general set of SMEFT operators. For example, Ref.~\cite{deBlas:2022hdk} fitted electroweak precision observables (EWPO) under the assumption of flavor universality, finding that the $W$-boson mass anomaly requires non-zero values of various dimension-six SMEFT Wilson coefficients. An important ingredient in these fits is the decay of the muon that enters in the determination of the Fermi constant. 
The hadronic counterparts of muon decay provide complementary probes of the Fermi constant in the form of $\beta$-decay processes of the neutron and atomic nuclei~\cite{Falkowski:2020pma}, and semileptonic meson decays~\cite{Gonzalez-Alonso:2016etj}.
Although the combination of EWPO and $\beta$-decay processes has been discussed in the literature before \cite{Crivellin:2021njn,Crivellin:2020ebi}, very few of the new analyses of the $W$-boson mass \cite{Blennow:2022yfm,Bagnaschi:2022whn} included the low-energy data.

In this section, we argue that not including constraints from these observables in global fits generally leads to too large deviations in first-row CKM unitarity, much larger than the mild tension shown in state-of-the-art determinations. We also investigate what solutions to the $W$-boson mass anomaly survive after including $\beta$-decay constraints \cite{Cirigliano:2022qdm}.

\subsection{Mass of the $W$-boson in SMEFT}

We adopt the parameterization of SMEFT at dimension-six in the \textit{Warsaw} basis~\cite{Weinberg:1979sa, Buchmuller:1985jz,Grzadkowski:2010es}.
\begin{equation}
	\mathcal{L}_\text{SMEFT}^{\text{dim-6}} =
	\mathcal{L}_\text{SM} +
	\sum_{i} C_i\mathcal{O}^{\text{dim-6}}_i \, ,
	\label{eq:LSMEFT}
\end{equation}
where $C_i$ are Wilson coefficients of mass dimension $-2$.

Calculated at linear order in SMEFT, the shift to $W$-boson mass from the SM prediction due to dimension-six operators is given by \cite{Berthier:2015oma,Bjorn:2016zlr}
\begin{equation}
	\frac{\delta m_W^2}{m_W^2} = 
	v^2 \ \frac{s_w c_w}{s_w^2 - c_w^2} 
	\left[ 2 \, C_{HWB} + \frac{c_w}{2 s_w} \, C_{HD} + 
	\frac{s_w}{c_w} \left( 2 \, C_{Hl}^{(3)} - C_{ll} \right) \right] \,,
	\label{eq:mW}
\end{equation}
where $v\simeq 246$ GeV is the vacuum expectation value of the Higgs field, $s_w = \sin{\theta_w}$ and $c_w = \cos{\theta_w}$.
The Weinberg angle $\theta_w$ is fixed by the electroweak input parameters $\{ G_F, m_Z, \alpha_{EW} \}$ \cite{Brivio:2021yjb}. Here we define $\delta m_W^2 = m_W^2(\mathrm{SMEFT}) -m_W^2(\mathrm{SM})$.

The mass of the $W$-boson receives corrections from four Wilson coefficients, namely $C_{HWB}$, $C_{HD}$, $C_{Hl}^{(3)}$, and $C_{ll}$. For the corresponding operators, see Tab.~\ref{tab:SMEFT_Ops}. $C_{HWB}$ and $C_{HD}$ are related to the oblique parameters $S$ and $T$ \cite{PhysRevLett.65.964}. They have been thoroughly studied for constraining 'universal' theories \cite{Barbieri:2004qk,Wells:2015uba} with electroweak precision observables as well as in light of the $W$-boson mass anomaly~\cite{Lu:2022bgw,Strumia:2022qkt,Paul:2022dds,Asadi:2022xiy}. The linear combination of Wilson coefficients shown in Eq.~\eqref{eq:mW}, $\left( 2 \, C_{Hl}^{(3)} - C_{ll} \right)$, is related to the shift to Fermi constant.

\begin{table*}[t]
	\centering
	\def\arraystretch{1.5}
	\begin{tabular}{|c|c|}
		\hline
		${\cal O}_{HWB}$ & $H^\dagger \tau^I H\, W^I_{\mu\nu} B^{\mu\nu}$ \\
		\hline
		${\cal O}_{HD}$ & $\big| H^\dagger D_\mu H \big|^2$ \\
		\hline
		${\cal O}_{Hl}^{(3)}$ & $\left(H^\dagger i \DLR^I_{\mu} H \right) \left(\bar l_p \tau^I \gamma^\mu l_r \right)$ \\
		\hline
		${\cal O}_{Hq}^{(3)}$ & $\left(H^\dagger i \DLR^I_{\mu} H \right) \left(\bar q_p \tau^I \gamma^\mu q_r \right)$ \\
		\hline
		${\cal O}_{ll}$ & $\left(\bar l_p \gamma_\mu l_r \right) \left(\bar l_s \gamma^\mu l_t \right)$ \\
		\hline
		${\cal O}_{lq}^{(3)}$ & $\left(\bar l_p \tau^I \gamma_\mu l_r \right) \left(\bar q_s \tau^I \gamma^\mu q_t \right)$ \\
		\hline
	\end{tabular}
	\caption{List of the most relevant SMEFT dimension-six operators that are involved in this analysis.}
	\label{tab:SMEFT_Ops}
\end{table*}

\subsection{EWPO fits and CKM unitarity}

Under the assumption of flavor universality, 10 operators affect the EWPO at tree level, but only 8 linear combinations can be determined by data \cite{deBlas:2022hdk}. Following Ref.~\cite{deBlas:2022hdk}, these linear combination are written with $\hat C_i$ notation and given by $\hat C_{Hf}^{(1)} = C_{Hf}^{(1)} - (Y_f / 2) C_{HD}$, where $f$ runs over left-handed lepton and quark doublets and right-handed quark and lepton singlets, and $\hat C_{Hf}^{(3)} = C_{Hf}^{(3)} + (c_w / s_w) C_{HWB} + (c_w^2 / 4 s_w^2) C_{HD}$ where $f$ denotes left-handed lepton and quark doublets, and $\hat C_{ll} = (C_{ll})_{1221}$. Here $Y_f$ is the hypercharge of the fermion $f$. 

Ref.~\cite{deBlas:2022hdk} reported the results of their fits including the correlation matrix from which we can reconstruct the $\chi^2$. 
We obtain very similar results by constructing a $\chi^2$ using the SM and SMEFT contributions from Refs.\ \cite{deBlas:2021wap} and \cite{Berthier:2015oma}, respectively, and the experimental results of Refs.\ \cite{ALEPH:2005ab,SLD:2000jop,Janot:2019oyi,Zyla:2020zbs}.
For concreteness we use the `standard average' results of Ref.\ \cite{deBlas:2022hdk}, but our point would hold for the `conservative average' as well. 
To investigate the consequences of CKM unitarity on the fit, we will assume the flavor structures of the operators follow Minimal Flavor Violation (MFV) \cite{Chivukula:1987py, DAmbrosio:2002vsn}. That is, we assume the operators are invariant under a $U(3)_q\times U(3)_u\times U(3)_d\times U(3)_l\times U(3)_e$ flavor symmetry.
In addition, we slightly change the operator basis and trade the Wilson coefficient $\hat C_{ll}$  for the linear combination
\begin{equation}
 C_{\Delta}= 2 \left[ C_{Hq}^{(3)} - C_{Hl}^{(3)} + \hat C_{ll} \right]\,,
\end{equation}
which allows for a more direct comparison to low-energy measurements. In principle one could trade $C_\Delta$ for one of the $\hat C_{Hf}^{(3)}$ coefficients instead of $\hat C_{ll}$, but this would not change the determination of $C_\Delta$.

We then refit the Wilson coefficients to the EWPO and obtain the results in the second column of Table~\ref{tab:SMEFT1}. In particular, we obtain
\begin{equation}
\label{deBlas}
C_{\Delta} = -\left(0.19\pm 0.09\right)\, {\rm TeV}^{-2}\,.
\end{equation}

This combination of Wilson coefficients contributes to the violation of unitarity in the first row of the CKM matrix tracked by $\Delta_{\mathrm{CKM}} \equiv |V_{ud}|^2+|V_{us}|^2-1$, 
where we neglected the tiny $ |V_{ub}|^2$ corrections. Within the MFV assumption, we can write \cite{Cirigliano:2009wk} 
\begin{equation}
\Delta_{\mathrm{CKM}} =v^2 \left[ C_{\Delta} - 2\,C_{lq}^{(3)} \right] \,.
\end{equation}
The $C_{lq}^{(3)}$ operator that appears here does not affect EWPO and does not play a role in the fit of Ref.~\cite{deBlas:2022hdk}. If one assumes this coefficient to be zero, Eq.~\eqref{deBlas} causes a shift 
\begin{equation}\label{CKMfit}
\Delta^{\mathrm{EWfit}}_{\mathrm{CKM}} =-(0.012\pm0.005)\,,
\end{equation}
implying large, percent-level, deviations from CKM unitarity. We note that, to a lesser extent, large values of $C_\Delta$ and $\Delta_{\rm CKM}$ were already preferred by fits to EWPO before the recent CDF determination of $m_W$. Using an older average of determinations of $m_W =80.379\pm0.012$ GeV \cite{deBlas:2021wap}, we find $C_\Delta = -0.15\pm0.09$ TeV$^{-2}$ and $\Delta_{\rm CKM} = -(0.9\pm 0.5)\cdot10^{-2}$ (assuming $C_{lq}^{(3)}=0$).

Based on up-to-date theoretical predictions for $0^+ \rightarrow 0^+$ transitions and Kaon decays \cite{Cirigliano:2008wn, FlaviaNetWorkingGrouponKaonDecays:2010lot,Seng:2018yzq,Czarnecki:2019mwq,Hardy:2020qwl,Aoki:2021kgd,Seng:2021nar,Seng:2022wcw}, the PDG average indicates that unitarity is indeed violated by a bit more than two standard deviations  
\cite{Zyla:2020zbs}
\begin{equation}
\label{PDG}
\Delta_{\mathrm{CKM}}  = -0.0015(7)\,,
\end{equation}
but in much smaller amounts than predicted by Eq.~\eqref{CKMfit}. 
This exercise shows that global fits to EWPO and the W mass anomaly 
which assume MFV and $C_{lq}^{(3)}=0$, but include BSM physics beyond the oblique parameters S and T, such as the one of Ref.~\cite{deBlas:2022hdk}, 
are disfavored by $\beta$-decay data. While we did not repeat the fits of Refs.~\cite{Bagnaschi:2022whn, Balkin:2022glu}, the central values of their Wilson coefficients also indicate a negative percent-level shift to $\Delta_{\mathrm{CKM}}$, consistent with Eq.~\eqref{CKMfit}. 
 
\begin{figure}[!t]
\begin{center}
\includegraphics[trim={1.2cm 0 0 0},clip, width=0.8\textwidth]{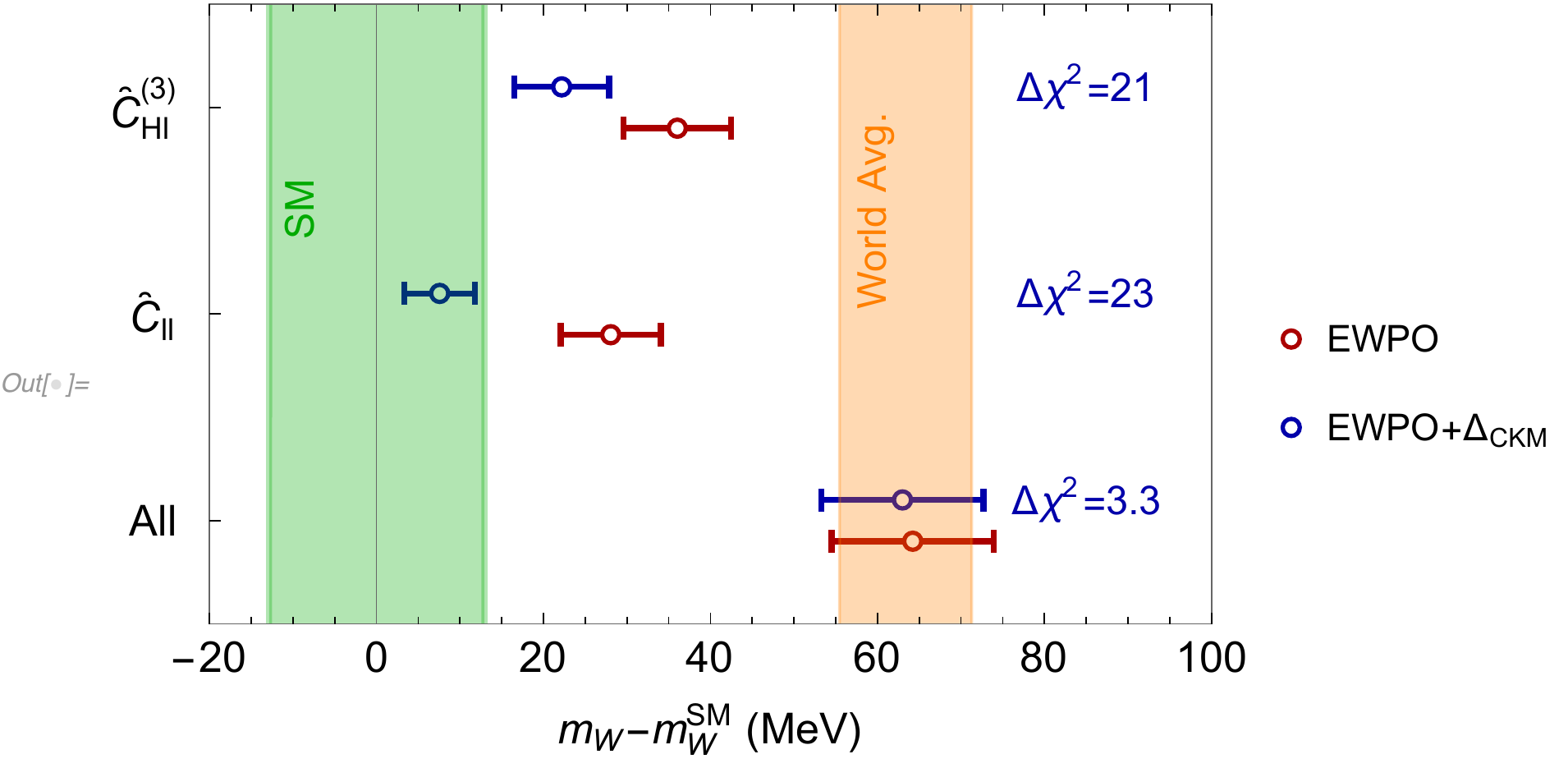}
\end{center}
\vspace{-0.5cm}
\caption{The resulting values of $\Delta m_W=m_W-m_W^{\rm SM}$ when turning on $\hat C_{Hl}^{(3)}$, $ \hat C_{ll}$, and all Wilson coefficients that are probed by EWPO. The red bars indicated the predicted $\Delta m_W$ from the EWPO fit, while the blue bars show the resulting $\Delta m_W$ after inclusion of $\Delta_{\mathrm{CKM}}$. The shown values of $\Delta \chi^2$, denote the differences in the minimum $\chi^2$ between the blue and red points. The SM prediction and world average, taken from Ref.\ \cite{deBlas:2022hdk}, are depicted by the green and orange bands, respectively. 
}\label{fig:mw}
\end{figure}

Indeed, combining the EWPO with $\Delta_{\rm CKM}$, we find that the minimum $\chi^2$ increases by $3.3$ and Wilson coefficients are shifted, as shown in Tab.~\ref{tab:SMEFT1}. 
Again this shows that the Cabibbo universality test has a significant impact and should 
be included in EWPO analyses of the $W$-boson mass anomaly. These statements are illustrated in Fig.~\ref{fig:mw}, 
which shows the values of 
$\Delta m_W=m_W-m_W^{\rm SM}$
obtained by fitting EWPO alone or EWPO {\it and} $\Delta_{\mathrm{CKM}} $
for two single-operator scenarios and the global analysis 
involving all operators.

\begin{table*}[t]
    \centering
    \begin{tabular}{c|c|c}
 \hline
 & Result & Result with CKM \\ 
 \hline
$\hat{C}_{\varphi l}^{(1)}$ & $ -0.007 \pm 0.011 $ &$ -0.013 \pm 0.009 $     \\ 
$\hat{C}_{\varphi l}^{(3)}$ & $ -0.042 \pm 0.015 $ & $-0.034\pm 0.014  $   \\ 
$\hat{C}_{\varphi e}$       & $ -0.017 \pm 0.009 $  &  $-0.021 \pm 0.009 $   \\
$\hat{C}_{\varphi q}^{(1)}$ & $ -0.0181\pm 0.044 $ &$  -0.048\pm 0.04   $   \\
$\hat{C}_{\varphi q}^{(3)}$ & $ -0.114 \pm 0.043 $  &$ -0.041\pm 0.015   $   \\
$\hat{C}_{\varphi u}$       & $ \phantom{+}0.086 \pm 0.154 $       & $-0.12\pm0.11 $    \\
$\hat{C}_{\varphi d}$       & $ -0.626 \pm 0.248 $   &  $-0.38  \pm 0.22 $   \\
$C_{\Delta}$          & $  -0.19 \pm 0.09 $     &  $ -0.027 \pm 0.011 $ \\ \hline
    \end{tabular}
    \caption{Results from the dimension-six SMEFT fit of Ref.~\cite{deBlas:2022hdk}, before and after the inclusion of $\Delta_{\mathrm{CKM}}$. All Wilson coefficients are given in units of TeV$^{-2}$. }
    \label{tab:SMEFT1}
\end{table*}

\begin{figure}[!t]
\begin{center}
\includegraphics[width=0.6\textwidth]{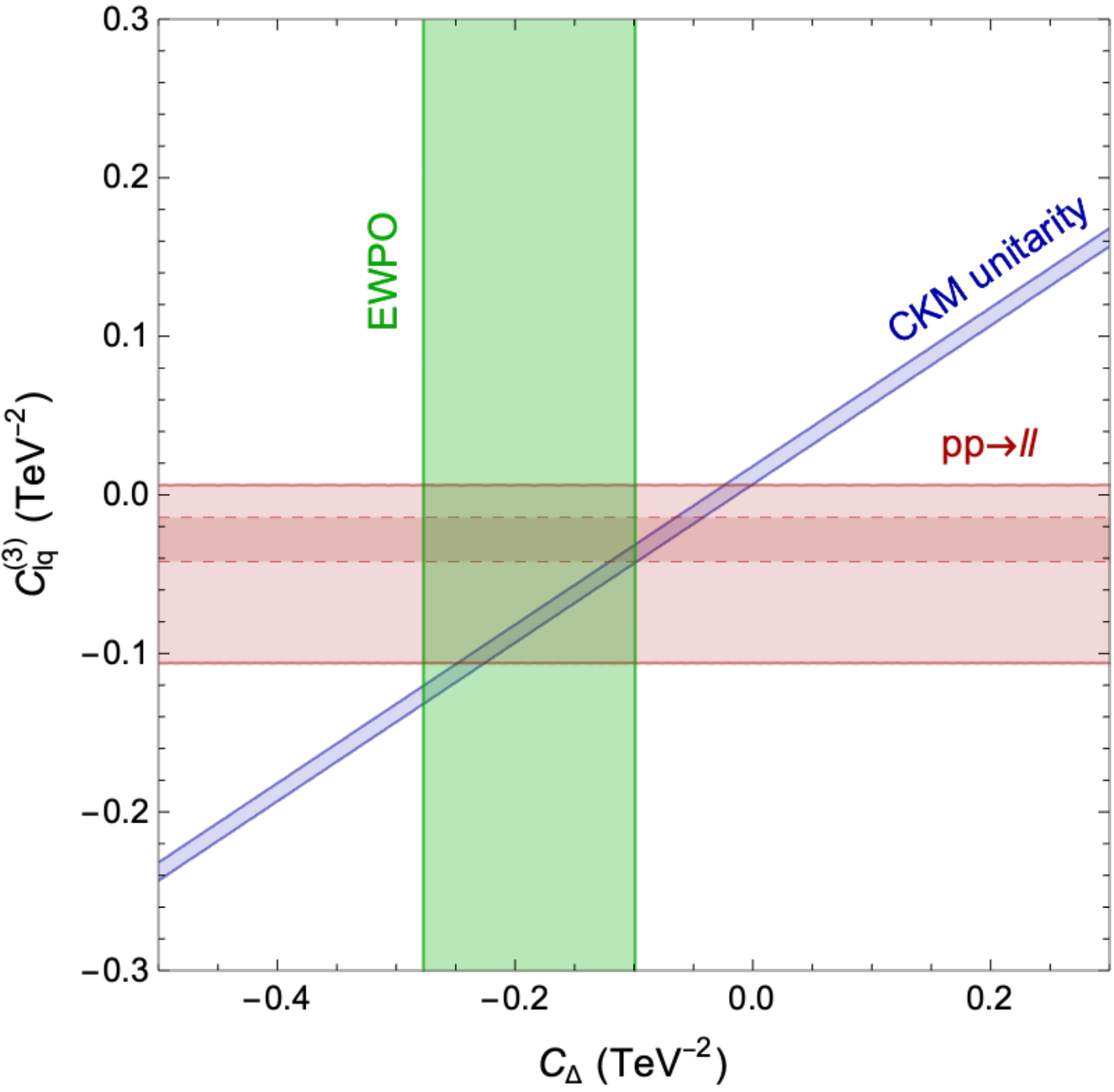}
\end{center}
\vspace{-0.5cm}
\caption{The $1\sigma$ constraints from EWPO in green, a global (single-coupling) analysis of LHC measurements in (dashed) red, and low-energy beta decays in blue. 
}\label{fig:plot}
\end{figure}

Another way to proceed is to effectively decouple the CKM unitarity constraint from EWPO by 
letting $C_{lq}^{(3)}\neq0$, which is 
consistent with the MFV approach. The $\Delta_{\mathrm{CKM}}$ observable is then accounted for by a nonzero value 
\begin{equation}
C_{lq}^{(3)} = -(0.082 \pm 0.045)\, {\rm TeV}^{-2}\,,
\end{equation}
while the values of the other Wilson coefficients return to their original value given in the second column of Table~\ref{tab:SMEFT1}. However, care must be taken that such values of $C_{lq}^{(3)}$ are not excluded by LHC constraints \cite{Bhattacharya:2011qm,Cirigliano:2012ab,Alioli:2018ljm,Panico:2021vav,Kim:2022amu,Greljo:2021kvv,Iranipour:2022iak}. 
In particular, Ref.\ \cite{Boughezal:2021tih} analysed
8 TeV $pp\to ll$ data  from \cite{ATLAS:2016gic} in the SMEFT at dimension-8.
Limiting the analysis to MFV dimension-six operators, we find 
\bea
C_{lq}^{(3)}&=& -(0.028\pm 0.028)\, {\rm TeV}^{-2}\quad ({\rm Single\, coupling},\quad 95\%\, {\rm C.L.})\,,\nonumber\\
C_{lq}^{(3)}&=& -(0.05\pm 0.1)\, {\rm TeV}^{-2}\quad\quad \,\,\,({\rm Global\, fit},\quad 95\%\, {\rm C.L.})\,,\label{eq:lhc}
\eea
when in the first line only $C_{lq}^{(3)}$ is turned on,
while in second line seven operators were turned on: 
$C_{lq}^{(1)}$, $C_{lq}^{(3)}$,
$C_{qe}$, $C_{lu}$, $C_{ld}$, $C_{eu}$, and $C_{ed}$.

The resulting constraints from EWPO, $\Delta_{\mathrm{CKM}}$, and the LHC are shown in Fig.\ \ref{fig:plot}. As mentioned above, a simultaneous explanation of $m_W$ and $\Delta_{\mathrm{CKM}}$ requires a nonzero value of $C_{lq}^{(3)}$, which implies effects in collider processes. The single-coupling bound from $pp\to ll$ in Eq.\ \eqref{eq:lhc} is already close to excluding the overlap of the EWPO and $\Delta_{\mathrm{CKM}}$ regions, while a global fit allows for somewhat more room. Nevertheless, should the current discrepancy in the EWPO fit hold, the preference for a nonzero $C_{lq}^{(3)}$ could be tested by existing 13 TeV $pp \rightarrow l l$ \cite{ATLAS:2020yat} and $pp \rightarrow l \nu$ data \cite{ATLAS:2019lsy}, and, in the future, at the HL-LHC.

\subsection{Flavor assumptions}

In this section we choose to only consider the case of $U(3)^5$ flavor invariance for the SMEFT operators. Several global fits in the literature adopted the same flavor assumption \cite{deBlas:2022hdk,Bagnaschi:2022whn}, so that we can meaningfully compare our results.
These flavor assumptions are ideal for describing flavor-blind BSM physics and, in practice, drastically reduce the number of Wilson coefficients entering the fit.
Relaxing this hypothesis has several implications: 
first, one should consider flavor non-universal Wilson coefficients for the usual set of operators in EWPO fits, see e.g.\ Ref.\ \cite{Balkin:2022glu}; second, when including semileptonic low-energy processes, one should 
extend the operator set to include operators with more general Lorentz structures than $O_{lq}^{(3)}$, such as right-handed currents \cite{Alioli:2017ces}, that  provide additional ways to decouple the 
$\Delta_{\rm CKM}$ constraint from the W mass.  
The inclusion of flavor non-universal Wilson coefficients~\cite{Balkin:2022glu} in the EWPO fit 
has been shown to provide a good solution to the W mass anomaly. 
In this context, addressing the role of CKM unitarity constraints and lepton-flavor universality tests in meson decays is an interesting direction. 

\section{Acknowledgements} J.M. gratefully acknowledges the financial support by ICREA under the ICREA Academia programme and from the Pauli Center (Zurich) and thanks to
the Physics Department of University of Zurich for the hospitality.
J.M.  received also financial support from Spanish Ministry of Science, Innovation and Universities (project PID2020-112965GB-I00) and from the Research Grant Agency of the Government of Catalonia (project SGR 1069). T.T. is supported by the Physics Department of University of Siegen.
A.M. gratefully acknowledges the financial support from the Swiss National Science Foundation (SNF) under project P400P2\_191121. AVK thanks colleagues of the CDF Collaboration and Fermilab and 
gratefully acknowledges the financial support from the respective funding agencies.
The computations for the BMWc determination of $a_\mu$ were performed on JUQUEEN, JURECA,
JUWELS and QPACE at Forschungszentrum J\"ulich, on SuperMUC and SuperMUC-NG at
Leibniz Supercomputing Centre in M\"unchen, on Hazel Hen and HAWK at the High
Performance Computing Center in Stuttgart, on Turing and Jean Zay at CNRS'
IDRIS, on Joliot-Curie at CEA's TGCC, on Marconi in Roma and on GPU clusters in
Wuppertal and Budapest. L.V. thanks the Gauss Centre for Supercomputing, PRACE and
GENCI (grant 52275) for awarding computer time for this project on these machines. This
project was partially funded by the DFG grant SFB/TR55, by the BMBF Grant No.
05P18PXFCA, by the Hungarian National Research, Development and Innovation
Office grant KKP126769 and by the Excellence Initiative of Aix-Marseille
University - A*MIDEX, a French “Investissements d’Avenir” program, through
grants AMX-18-ACE-005, AMX-19-IET-008 - IPhU and ANR-11-LABX-0060. L.V. thanks all his colleges in the BMW collaboration.

\providecommand{\href}[2]{#2}\begingroup\raggedright

\end{document}